\documentclass[11pt]{article}
\usepackage{amsmath} % it subsumes: amstext, amsbsy, amsopn
\usepackage[dvips]{graphics}
\usepackage{latexsym}
\usepackage{psfig}
\usepackage{amscd}     \usepackage{amsxtra}
\usepackage{upref}     \usepackage{amsthm}
\usepackage{amssymb}   %\usepackage{mathrsfs}
\usepackage{amsfonts}
\begin{document}

\title{\bf Nonlinear Connections and Description of Photon-like Objects}
%\author{{\bf Stoil Donev}\footnote{e-mail:

\author{
{\bf Stoil Donev$^1$\footnote{e-mail:
 sdonev@inrne.bas.bg} \ , Maria Tashkova$^1$}
\noindent
\\ (1) Institute for Nuclear Research and Nuclear Energy,\\
Bulg.Acad.Sci., 1784 Sofia, blvd.Tzarigradsko chaussee 72, Bulgaria}
\date{}
\maketitle

\begin{abstract}
This paper aims to present a general idea for description of spatially finite physical objects with
a consistent nontrivial translational-rotational dynamical structure and evolution as a whole,
making use of the mathematical concepts and structures connected with the Frobenius
integrability/nonintegrability theorems given in terms of distributions
on manifolds with corresponding curvature defined by the Nijenhuis operator.
The idea is based on consideration of {\it nonintegrable} subdistributions of
some appropriate completely integrable distribution (differential system) on a
manifold and then to make use of the corresponding curvatures as generators of
measures of interaction, i.e. of energy-momentum exchange among the physical
subsystems mathematically represented by the nonintegrable subdistributions.
The concept of photon-like object is introduced and description of such objects
in these terms is given.

 \end{abstract}

\section{Introduction} At the very dawn of the 20th century Planck (Planck 1901) proposed and a
little bit later Einstein (Einstien 1905) appropriately used the well known and widely used through
the whole last century simple formula $E=h\nu$, $h=const>0$.  This formula marked the beginning of
a new era and became a real symbol of the physical science during the following years. According to
the Einstein's interpretation it gives the full energy $E$ of {\it really existing} light quanta of
frequency $\nu=const$, and in this way a new understanding of the nature of the electromagnetic
field was introduced:  the field has structure which contradicts the description given by Maxwell
vacuum equations.  After De Broglie's (De Broglie 1923) suggestion for the particle-wave nature of
the electron obeying the same energy-frequency relation, one could read Planck's formula in the
following way:  {\it there are physical objects in Nature the very existence of which is strongly
connected to some periodic (with time period $T=1/\nu$) process of intrinsic for the object nature
and such that the Lorentz invariant product $ET$ is equal to $h$}. Such a reading should suggest
that these objects do NOT admit point-like approximation since the relativity principle for free
point particles requires straight-line uniform motion, hence, no periodicity should be allowed.

Although the great (from pragmatic point of view) achievements of the developed theoretical
approach, known as {\it quantum theory}, the great challenge to build an adequate description of
individual representatives of these objects, especially of light quanta called by Lewis {\it
photons} (Lewis 1926) is still to be appropriately met since  the efforts made in this direction,
we have to admit, still have not brought satisfactory results. Recall that Einstein in his late
years recognizes (Speziali 1972) that "the whole fifty years of conscious brooding have not brought
me nearer to the answer to the question "what are light quanta", and now, half a century later,
theoretical physics still needs progress to present a satisfactory answer to the question "what is
a photon". We consider the corresponding theoretically directed efforts as necessary and even {\it
urgent} in view of the growing amount of definite experimental skills in manipulation with
individual photons, in particular, in connection with the experimental advancement in the "quantum
computer" project.  The dominating modern theoretical view on microobjects is based on the notions
and concepts of quantum field theory (QFT) where the structure of the photon (as well as of any
other microobject) is accounted for mainly through the so called {\it structural function}, and
highly expensive and delicate collision experiments are planned and carried out namely in the frame
of these concepts and methods (see the 'PHOTON' Conferences Proceedings, some recent review papers:
Dainton 2000; Stumpf, Borne 2001; Godbole 2003; Nisius 2001).  Going not in details we just note a
special feature of this QFT approach: if the study of a microobject leads to conclusion that it has
structure, i.e. it is not point-like, then the corresponding constituents of this structure are
considered as point-like, so the point-likeness stays in the theory just in a lower level.

In this paper we follow another approach based on the assumption that the description of the
available (most probably NOT arbitrary) spatial structure of photon-like objects can be made by
{\it continuous finite/localized} functions of the three space variables.  The difficulties met in
this approach consist mainly, in our view, in finding adequate enough mathematical objects and
solving appropriate PDE.  The lack of sufficiently reliable corresponding information made us look
into the problem from as general as possible point of view on the basis of those properties of
photon-like objects which may be considered as most undoubtedly trustful, and
in some sense, {\it identifying}. The analysis made suggested that such a
property seems to be the {\it the available translational-rotational
dynamical structure}, so we shall focus on this property in order to see what useful for
our purpose suggestions could be deduced and what appropriate structures could
be constructed. All these suggestions and structures should be the building
material for a step-by-step creation of a {\it self-consistent} system. From
physical point of view this should mean that the corresponding properties may combine to express a
dynamical harmony in the inter-existence of appropriately defined subsystems of
a finite and time stable larger physical system. (for another approach based on
slight modification of Maxwell equations see Funaro,D., arXiv:physics/0505068)

The plan of this paper is the following. In Sec.2 we introduce and comment the
concept of {\it photon-like object}. In Sec.3 we recall some basic facts from
Frobenius integrability theory, then we consider its possibilities to describe
interaction between/among subsystems, mathematically represented by
non-integrable subdistributions of an integrable distribution. In Sec.4 we
introduce and comsider the concept of nonlinear connection and deduce some
important from our point of view relations. In Secs.5,6 we consider
electromagnetic pfoton-like objects. In Sec.7 we interpret geometrically the
translational-rotational consistency and obtain another look at the equations
of motion obtained in Sec.5. In the concluding Sec.8 we give a short overview
of the results obtained.

\section{The notion of photon-like object} We begin with the notice that any notion of a physical
object must unify two kinds of properties of the object considered: {\it identifying}
and {\it kinematical}. The identifying properties being represented by
quantities and relations, stay unchanged throughout the existence, i.e.
throughout the time-evolution, of the object, they represent all the intrinsic
structure and relations.  The kinematical properties describe those changes,
called {\it admissible}, which do NOT lead to destruction of the object, i.e.
to the destruction of any of the identifying properties. Correspondingly,
physics introduces two kinds of quantities and relations, identifying and
kinematical. From theoretical point of view the more important quantities used
turn out to be the {\it dynamical} quantities which, as a
rule, are functions of the identifying and kinematical ones, and the joint
relations they satisfy represent the necessary interelations between them in
order this object to survive under external influence. This view suggests to
introduce the following notion of Photon-like object(s) (we shall use the
abbreviation "PhLO" for "Photon-like object(s)"):

\begin{center} {\bf PhLO are real massless time-stable physical objects with a consistent
translational-rotational dynamical structure}. \end{center}

We give now some explanatory comments, beginning with the term {\it real}. {\bf
First} we emphasize that this term means that we consider PhLO as {\it really}
existing {\it physical} objects, not as appropriate and helpful but imaginary
(theoretical) entities.  Accordingly, PhLO {\bf necessarily carry
energy-momentum}, otherwise, they could hardly be detected.  {\bf Second}, PhLO
can undoubtedly be {\it created} and {\it destroyed}, so, no point-like and
infinite models are reasonable: point-like objects are assumed to have no
structure, so they can not be destroyed since there is no available structure
to be destroyed; creation of infinite physical objects (e.g. plane waves)
requires infinite quantity of energy to be transformed from one kind to another
for finite time-period, which seems also unreasonable. Accordingly, PhLO are
{\it spatially finite} and have to be modeled like such ones, which is the only
possibility to be consistent with their "created-destroyed" nature. It seems
hardly reasonable to believe that PhLO can not be created and destroyed, and
that spatially infinite and indestructible physical objects may exist at all.
{\bf Third}, "spatially finite" implies that PhLO may carry only {\it finite
values} of physical (conservative or non-conservative) quantities.  In
particular, the most universal physical quantity seems to be the
energy-momentum, so the model must allow finite integral values of
energy-momentum to be carried by the corresponding solutions. {\bf Fourth},
"spatially finite" means also that PhLO {\it propagate}, i.e.  they do not
"move" like classical particles along trajectories, therefore, partial
differential equations should be used to describe their evolution in time.

The term "{\bf massless}" characterizes physically the way of propagation in
terms of appropriate dynamical quantities: the {\it integral} energy $E$ and
{\it integral} momentum $p$ of a PhLO should satisfy the relation $E=cp$, where
$c$ is the speed of light in vacuum, and in relativistic terms this means that
their integral energy-momentum vector {\it must be isotropic}, i.e. it must
have zero module with respect to Lorentz-Minkowski (pseudo)metric in
$\mathbb{R}^4$. If the object considered has spatial and time-stable structure,
so that the translational velocity of every point where the corresponding field
functions are different from zero must be equal to $c$, we have in fact null
direction in the space-time {\it intrinsically determined} by a PhLO. Such a
direction is formally defined by a null vector field $X,X^2=0$. The integral
trajectories of this vector field are isotropic (or null) straight lines as is
traditionally assumed in physics. It follows that with every PhLO a null
direction is {\it necessarily} associated, so, canonical coordinates
$(x^1,x^2,x^3,x^4)=(x,y,z,\xi=ct)$ on $\mathbb{R}^4$ may be chosen such that in
the corresponding coordinate frame $X$ to have only two non-zero components of
magnitude $1$: $X^\mu=(0,0,-\varepsilon, 1)$, where $\varepsilon=\pm 1$
accounts for the two directions along the coordinate $z$ (further such a
coordinate system will be called $X$-adapted and will be of main usage). Our
PhLO propagates as a whole along the $X$-direction, so the corresponding
energy-momentum tensor $T_{\mu\nu}$ of the model must satisfy the corresponding
{\it local isotropy (null) condition}, namely,  $T_{\mu\nu}T^{\mu\nu}=0$
(summation over the repeated indices is throughout used).

The term "{\bf translational-rotational}" means that besides translational component along
$X$, the propagation necessarily demonstrates some rotational (in the general
sense of this concept) component in such a way that {\it both components exist
simultaneously and consistently}. It seems reasonable to expect that such kind
of behavior should be consistent only with some distinguished spatial shapes.
Moreover, if the Planck relation $E=h\nu$ must be respected throughout the
evolution, the rotational component of propagation should have {\it
time-periodical} nature with time period $T=\nu^{-1}=h/E=const$, and one of the
two possible, {\it left} or {\it right}, orientations. It seems reasonable also
to expect periodicity in the spatial shape of PhLO, which somehow to be related
to the time periodicity.

The term "{\bf dynamical structure}" means that the propagation is supposed to be necessarily
accompanied by an {\it internal energy-momentum redistribution}, which may be considered in the
model as energy-momentum exchange between (or among) some appropriately defined subsystems.  It
could also mean that PhLO live in a dynamical harmony with the outside world, i.e.  {\it any
outside directed energy-momentum flow should be accompanied by a parallel inside directed
energy-momentum flow}.

Finally, note that if the time periodicity and the spatial periodicity
should be consistently related somehow, the simplest integral feature of such
consistency would seem like this: the spatial size along the translational
component of propagation $\lambda$ is equal to $cT$: $\lambda=cT$, where $\lambda$ is
some finite positive characteristic constant of the corresponding solution.
This would mean that every individual PhLO determines its own length/time
scale.
\vskip0.3cm
We are going now to formulate shortly the basic idea, i.e. the basic
mathemetical identification, inside which this study will be carried out.

\section{Curvature of Distributions and Physical Interaction}

Any physical system with a dynamical structure is characterized by some
internal energy-momentum redistributions, i.e. energy-momentum fluxes, during
evolution. Any system of energy-momentum fluxes (as well as fluxes of other
interesting for the case physical quantities subject to change during
evolution, but we limit ourselves just to energy-momentum fluxes here) can be
considered mathematically as generated by some system of vector fields. A {\it
physically isolated, consistent} and {\it interelated time-stable} system of
energy-momentum fluxes can be considered to correspond directly or indirectly
to a completely integrable distribution $\Delta$ of vector fields (or
differential system (Godbillon 1969)) according to the principle {\it some
local objects can generate integral object}. Every distribution on a manifold
defines its own curvature form (given further in the section). Let $\Delta_1$
and $\Delta_2$ be two distributions on the same manifold with corresponding
curvature forms $\Omega_1$ and $\Omega_2$, each of them carries couples of
vector fields inside their distributions outside $\Delta_1$ and $\Delta_2$
correspondingly, i.e. $\Omega_1(Y_1,Y_2)$ is out of $\Delta_1$ and
$\Omega_2(Z_1,Z_2)$ is out $\Delta_2$, where $(Y_1,Y_2)$ live in $\Delta_1$ and
$(Z_1,Z_2)$ live in $\Delta_2$. Let now $\Delta_1$ and $\Delta_2$ characterize
two interacting physical systems, or two interacting subsystems of a larger
physical system. It seems reasonable to assume as a workong tool the following
geometrization of the concept of local physical interaction: {\it two
distributions $\Delta_1$ and $\Delta_2$ on a manifold will be said to interact
infinitesimally (or locally) if at least one of the corresponding two curvature
forms $\Omega_1$/$\Omega_2$ takes values, or generates objects taking values,
respectively in $\Delta_2$/$\Delta_1$}.

The above geometric concept of {\it infinitesimal interaction} is motivated by
the fact that, in general, an integrable distribution $\Delta$ may contain
various {\it nonintegrable} subdistributions $\Delta_1, \Delta_2, \dots$ which
subdistributions may be associated physically with interacting subsytems of a
larger time stable physical system. Any physical interaction between 2
subsystems is necessarily accompanied with available energy-momentum exchange
between them, this could be understood mathematically as nonintegrability of
each of the two subdistributions of $\Delta$ and could be naturally measured
directly or indirectly by the corresponding curvatures. For example, if
$\Delta$ is an integrable 3-dimensional distribution spent by the vector fields
$(X_1,X_2,X_3)$ then we may have, in general, three non-integrable, i.e.
geometrically interacting, 2-dimensional subdistributions $(X_1,X_2),
(X_1,X_3), (X_2,X_3)$. Finally, some interaction with the outside world can be
described by curvatures of distributions (and their subdistributions) in which
elements from $\Delta$ and vector fields outside $\Delta$ are involved (such
processes will not be considered in this paper).

There are two basic ways to formalize the obove statements. The first one is
known as the {\bf Frobenius integrability approach}, and the second one (been
developed recently) is known as {\bf nonlinear connections} (Vacaru, S. et al.
2005). We consider briefly the first one here and then go in a more detail to
the nonlinear connections approach.

According to the Frobenius integrability theorem on a n-dimensional manifold $M^n$
(further all manifolds are assumed smooth and
finite dimensional and all objects defined on $M^n$ are also assumed smooth)
if the system of vector fields
$\Delta=\left[X_1(x),X_2(x),\dots, X_p(x)\right]$, $x\in M$, $1<p<n$, satisfies
$X_1(x)\wedge X_2(x)\wedge \dots ,\wedge\, X_p(x)\neq 0, \,x\in M$, then
$\Delta$ is completely integrable iff all Lie brackets $\left[X_i,X_j\right], \
i,j=1,2,\dots, p$ are representable linearly through the very $X_i,
i=1,2,\dots, p: \left[X_i,X_j\right]=C^k_{ij}X_k$, where $C^k_{ij}$ are
functions. Clearly, an easy way to find out if a distribution is
completely integrable is to check if the exterior products
\begin{equation}
[X_i,X_j]\wedge
X_1(x)\wedge X_2(x)\wedge \dots ,\wedge\, X_p(x), \,x\in M;\ \ \
i,j=1,2,\dots,p
\end{equation}
are identically zero. If this is not the case (which means
that at least one such Lie bracket "sticks out" of the distribution $\Delta$)
then the corresponding coefficients, which are multilinear combinations of the
components of the vector fields and their derivatives, represent the
corresponding curvatures. We note finally that if two subdistributions contain
at least one common vector field it seems naturally to expect interaction.

In the dual formulation of Frobenius theorem in terms of differential 1-forms
(i.e. Pfaff forms), having the distribution $\Delta$, we
look for $(n-p)$-Pfaff forms $(\alpha^1, \alpha^2, \dots, \alpha^{n-p}$), i.e. a
$(n-p)$-codistribution $\Delta^*$, such that $ \langle\alpha^m,X_j\rangle=0,\
\ \text{and} \ \ \alpha^1\wedge\alpha^2\wedge\dots \wedge\alpha^{n-p}\neq 0, $
$ m=1,2,\dots,n-p, \ \ j=1,2,\dots,p . $ Then the integrability of the
distribution $\Delta$ is equivalent to the requirements
\begin{equation}
\mathbf{d}\alpha^m\wedge\alpha^1\wedge\alpha^2\wedge\dots\wedge\alpha^{n-p} =0,\ \ \ m=1,2,\dots,
(n-p),
\end{equation}
where $\mathbf{d}$ is the exterior derivative.

Since the idea of curvature associated with, for example, an arbitrary
2-dimensional
distribution $(X,Y)$ is to find out if the Lie bracket $[X,Y]$ has components
along vector fields outside the 2-plane defined by $(X,Y)$, in our case
 we have to evaluate
the quantities $\langle\alpha^m,[X,Y]\rangle$, where all linearly independent
1-forms $\alpha^m$ annihilate
$(X,Y):\langle\alpha^m,X\rangle=\langle\alpha^m,Y\rangle=0$. In view of the
formula
$$ \mathbf{d}\alpha^m(X,Y)=X(\langle\alpha^m,Y\rangle)-
Y(\langle\alpha^m,X\rangle) -\langle\alpha^m,[X,Y]\rangle=
-\langle\alpha^m,[X,Y]\rangle
$$
we may introduce explicitly the curvature 2-form for the distribution
$\Delta(X)=(X_1,\dots,X_p)$. In
fact, if $\Delta(Y)=(Y_1,\dots,Y_{n-p})$ define a distribution which is
complimentary (in the sense of direct sum) to $\Delta(X)$ and
$\langle\alpha^m,X_i\rangle=0$,
$\langle\alpha^m,Y_n\rangle=\delta^m_n$, i.e. $(Y_1,\dots,Y_{n-p})$ and
$(\alpha^1, \dots, \alpha^{n-p})$ are dual bases, then the
corresponding curvature 2-form $\Omega_{\Delta(X)}$ should be defined by
\begin{equation}
\Omega_{\Delta(X)}=-\mathbf{d}\alpha^m\otimes Y_m, \ \ \text{since} \ \
\Omega_{\Delta(X)}(X_i,X_j)=-\mathbf{d}\alpha^m(X_i,X_j) Y_m=
\langle\alpha^m,[X_i,X_j]\rangle Y_m ,
\end{equation}
where it is meant here that
$\Omega_{\Delta(X)}$ is restricted to the distribution $(X_1,\dots,X_p)$.
Hence, if we call the distribution $(X_1,\dots,X_p)$ {\it horizontal} and the
complimentary distribution $(Y_1,\dots,Y_{n-p})$ {\it vertical}, then the
curvature 2-form acquires the status of {\it vertical bundle valued 2-form}. We
see that the curvature 2-form distinguishes those couples of vector fields
inside $\Delta(X)$ the Lie brackets of which define outside $\Delta(X)$
directed flows, and so, do not allowing to find integral manifold of
$\Delta(X)$. Clearly, the supposition here for dimensional complementarity of
the two distributions $\Delta(X)$ and $\Delta(Y)$ is not essential for the idea
of geometrical interaction, i.e. the distribution $\Delta(Y)\neq\Delta(X)$ may
be any other distribution on the same manifold with
dimension smaller than $(n-p)$, so that $m=1,2,\dots,q<(n-p)$ in general, the
important moment is that the two distributions (or subdistributions) can
"communicate" {\it differentially} through their curvature 2-forms.

Hence, from physical point of view, if the quantities
$\Omega_{\Delta(X)}(X_i,X_j)$ participate somehow in building the
components of the energy-momentum locally transferred from the system
$\Delta(X)$ to the system $\Delta(Y)$, then, naturally, we have to make use of
the quantities
$\Omega_{\Delta(Y)}(Y_m,Y_n)$ to build the components of
the energy-momentum transferred from $\Delta(Y)$ to $\Delta(X)$.
It deserves to note that it is possible a dynamical
equilibrium between the two systems $\Delta(Y)$ and
$\Delta(X)$ to exist: each system to gain as much energy-momentum as it
loses, and this to take place at every space-time point. On the other hand, the
restriction of $\Omega_{\Delta(X)}=-\mathbf{d}\alpha^m\otimes Y_m,
m=1,\dots,q$ to the system $\Delta(Y)$, i.e. the quantities
$\Omega_{\Delta(X)}(Y_m,Y_n)$, and the restriction of
$\Omega_{\Delta(Y)}=-\mathbf{d}\beta^i\otimes X_i, i=1,\dots,p,
\langle\beta^i, X_j\rangle=\delta^i_j,
\beta^1\wedge\dots\wedge\beta^p\neq 0,
\langle\beta^m,Y_i\rangle=0$, to $\Delta(X)$, i.e. the quantities
$\Omega_{\Delta(Y)}(X_i,X_j)$, acquire the sense of objects in terms of which
the local change of the corresponding energy-momentum, i.e. differences between
energy-momentum gains and losses, should be expressed. Therefore, if $W_{(X,Y)}$
denotes the energy-momentum transferred locally from $\Delta(X)$ to $\Delta(Y)$,
$W_{(Y,X)}$ denotes the energy-momentum transferred locally from $\Delta(Y)$ to
$\Delta(X)$, and $\delta W_{(X)}$ and $\delta W_{(Y)}$ denote respectively the
local energy-momentum changes of the two systems $\Delta(X)$ and $\Delta(Y)$,
then according to the local energy-momentum conservation law we can write
\[
\delta W_{(X)}=W_{(Y,X)}+W_{(X,Y)}, \ \
\delta W_{(Y)}=-(W_{(X,Y)}+W_{(Y,X)})=-\delta W_{(X)} .
\]
For the case of dynamical equilibrium we have
$W(X,Y)=-W(Y,X)=0$, so in such a case we obtain
\begin{equation}
\delta W_{(X)}=0,\ \ \ \delta
W_{(Y)}=0,\ \ \ W_{(Y,X)}+W_{(X,Y)}=0.
\end{equation}
As for how to build explicitly the
corresponding representatives of the energy-momentum fluxes, probably,
universal procedure can not be offered.
If, for example, the mathematical representative of the entire system
containing $\Delta(X)$ and $\Delta(Y)$ as subsystems, is a differential form
$G$, then the most simple procedure seems to be to "project" the curvature
components $\Omega_{\Delta(X)}(X_i,X_j)$ and $\Omega_{\Delta(Y)}(Y_m,Y_n)$, as
well as the components $\Omega_{\Delta(X)}(Y_i,Y_j)$ and
$\Omega_{\Delta(Y)}(X_m,X_n)$ on $G$, i.e. to consider the corresponding
interior products. For every special case, however, appropriate quantities
constructed out of the members of the introduced distributions and
co-distributions must be worked out.

\begin{center}
\section{Non-linear connections}
\end{center}
2.1.\underline{Projections}:
These are linear maps $P$ in a linear space $W^n$ sending all elements of $W^n$
to some subspace $P(W^n)\subset W^n$, so that $P\circ P=P$.
Let $(e_1,\dots,e_n)$ and $(\varepsilon^1,\dots,\varepsilon^n)$ be two dual
bases in $W^n$ and $(W^n)^*$, such that $P(W^n)$ is spent by
$(e_{p+1},\dots,e_n)$ and the dual to $P(W^n)$ is spent by
$(\varepsilon^{p+1}\dots,\varepsilon^n)$. The map $P$ is reduced to the
identity map in $P(W^n)$, so, it is given there by the tensor
$\varepsilon^a\otimes e_a$ , where $a=p+1,\dots,n$. The linear map $P$,
restricted to some other subspace $\mathbb{H}$ of $W^n$, such that
$\mathbb{H}\oplus Im(P)=W^n$ should be represented by some appropriate matrix
$N_{i}^{a}$ in the corresponding bases, so the map $P$ looks like in these
bases as follows:
\begin{equation}
P=\varepsilon^a\otimes e_a+(N_i)^a\varepsilon^i\times e_a,
\ \ i=1,\dots,p \ ; \ \ a=p+1,\dots,n \,.
\end{equation}

Let now $\phi$ and $\psi$ be two arbitrary linear maps, $\mathfrak{B}$ be
a bilinear map in $W^n$, and
$(\mathbf{x},\mathbf{y})$ be two arbitrary vectors in $W^n$. We consider the
expression
\[
\mathcal{A}(\mathfrak{B};\phi,\psi)(\mathbf{x},\mathbf{y})\equiv
\frac12\Big[\mathfrak{B}(\phi(\mathbf{x}),\psi(\mathbf{y}))+
\mathfrak{B}(\psi(\mathbf{x}),\phi(\mathbf{y}))+
\phi\circ\psi(\mathfrak{B}(\mathbf{x},\mathbf{y}))+
\psi\circ\phi(\mathfrak{B}(\mathbf{x},\mathbf{y}))
\]
\[
-\phi(\mathfrak{B}(\mathbf{x},\psi(\mathbf{y})))-
\phi(\mathfrak{B}(\psi(\mathbf{x}),\mathbf{y}))-
\psi(\mathfrak{B}(\mathbf{x},\phi(\mathbf{y})))-
\psi(\mathfrak{B}(\phi(\mathbf{x}),\mathbf{y}))\Big]\ \ .
\]
Assuming $\phi=\psi $ are projections denoted by $P$ this expression becomes
\[\mathcal{A}(\mathfrak{B};P)(\mathbf{x},\mathbf{y})\equiv
P(\mathfrak{B}(\mathbf{x},\mathbf{y}))+\mathfrak{B}(P(\mathbf{x}),P(\mathbf{y}))-
P(\mathfrak{B}(\mathbf{x},P(\mathbf{y})))-P(\mathfrak{B}(P(\mathbf{x}),\mathbf{y})) \ .
\]
Denoting the identity map of $W^n$ by $id$ and adding and subtracting
$P\Big[\mathfrak{B}\big(P(\mathbf{x}),P(\mathbf{y})\big)\Big]$, after some
elementary transformations we obtain
\[
\mathcal{A}(\mathfrak{B};P)(\mathbf{x},\mathbf{y})\equiv
P\Big[\mathfrak{B}\big[(id-P)(\mathbf{x}),(id-P)(\mathbf{y})\big]\Big]+
(id-P)\Big[\mathfrak{B}\big[P(\mathbf{x}),P(\mathbf{y}\big]\Big] .
\]
Recalling that $P$ and $(id-P)$ project on two subspaces of $W^n$,
the direct sum of which generates $W^n$,
and naming $P$ as {\it vertical} projection denoted by $V$, then
$(id-P)$, denoted by $H$, gets naturally the name {\it horizontal}
projection. So the above expression gets the final form of
\begin{equation}
\mathcal{A}(\mathfrak{B};P)(\mathbf{x},\mathbf{y})\equiv
V\Big[\mathfrak{B}\big[H(\mathbf{x}),H(\mathbf{y})\big]\Big]
+H\Big[\mathfrak{B}\big[V(\mathbf{x}),V(\mathbf{y})\big]\Big].
\end{equation}
Hence, the first term on the right measures the vertical component of the
$\mathfrak{B}$-image of the horizontal projections of
$(\mathbf{x},\mathbf{y})$, and the second term measures the horizontal
component of the $\mathfrak{B}$-image of the vertical projections of
$(\mathbf{x},\mathbf{y})$, which is in correspondence with the well known fact
that if $Ker(P)$ is the {\it kernal} space of $P$ and $Im(P)$ is the image
space of $P$ then the vector space $W^n$ is a direct sum of $Ker(P)$ and
$Im(P)$: $W^n=Ker(P)\oplus Im(P)$.

 We carry now this pure algebraic construction to the tangent bundle of a
smooth manifold $M^n$, where the above bilinear map $\mathfrak{B}$ will be
interpreted as the Lie bracket of vector fields, and the linear maps will be
just linear endomorphisms of the tangent/cotangent bundles of $M^n$.
Under these assumptions the image of the above initial
expression is called Nijenhuis bracket of the two linear endomorphisms $\Phi$
and $\Psi$, and is usually denoted by $[\Phi,\Psi]$. It has two important for
us properties: the first one is that it is linear with respect to the smooth
functions on the manifold, so, the Nijenhuis bracket allows, starting with two
$(1,1)$-tensors on $M^n$, to construct through differentiations one
$(2,1)$-tensor field being antisymmetric with respect to the covariant indices,
i.e. a 2-form that is valued in the tangent bundle of $M^n$; the second
property is that if $\Phi=\Psi$ then $[\Phi,\Phi]$ is not necessarily zero.

\vskip 0.3cm \noindent
 2.2 \underline{Nonlinear connections}

\noindent
Let now $(x^1,\dots,x^n)$ be any local coordinate system on our real manifold
$M^n$. We have the corresponding local frames $\{dx^1,\dots,dx^n\}$ and
$\{\partial_{x^1},\dots,\partial_{x^n}\}$. Let for each $x\in M$ we are given a
projection $P_x$ of the same constant rank $p$, i.e. $p$ does not depend on
$x$, in every tangent space $T_x(M)$. The space $Ker(P_x)\subset T_x(M)$ is
usually called $P$-{\it horizontal}, and the space $Im(P_x)\subset T_x(M)$ then
is called $P$-{\it vertical}. Thus, we have two distributions on $M$  the
direct sum of which gives the tangent bundle: $T(M)=Ker(P)\oplus Im(P)$.
The above result shows that each of these two distributions can be endowed with
corresponding 2-form, valued in the other distribution, and depending on some
binar operation in $TM^n$. As we mentioned the combination "Nijenhuis bracket
plus Lie bracket" leads to tensor field. Therefore, assuming that
the corresponding curvatures are defined by means of the combination
"Nijenhuis bracket of $P$ plus Lie bracket of vector fields"
we say that $P$ defines a nonlinear connection on $M$. Denoting by
$\mathcal{R}$ the so defined curvature 2-form of $Ker(P)$ and by
$\bar{\mathcal{R}}$ the analogically defined curvature 2-form of $Im(P)$, by
$V_{P}$ the restriction of $P$ to $Ker(P)$ and by $H_{P}$ the restriction of
$P$ to $Im(P)$, we can write
\begin{equation}
[P,P](X,Y)=\mathcal{R}(X,Y)+\bar{\mathcal{R}}(X,Y),
\end{equation}
where
$$
\mathcal{R}(X,Y)=V_{P}\big([H_{P}X,H_{P}Y]\big), \ \ \
\bar{\mathcal{R}}(X,Y)=H_{P}\big([V_{P}X,V_{P}Y]\big),
$$
$(X,Y)$ are any two
vector fields and the Lie bracket is denoted by $[ , ]$. Recalling the contents
of the preceding section, we can say that $\mathcal{R}(X,Y)\neq 0$ measures the
nonintegrability of the corresponding horizontal distribution, and
$\mathcal{\bar{R}}(X,Y)\neq 0$ measures the nonintegrability of the vertical
distribution.

If the vertical distribution is given before-hand and is completely integrable,
i.e. $\mathcal{\bar{R}}=0$,
then $\mathcal{R}(X,Y)$ is called {\it curvature} of the
nonlinear connection $P$ if there exist at least one couple of vector fields
$(X,Y)$ such that $\mathcal{R}(X,Y)\neq 0$.
\vskip 0.5 cm
\section{Photon-like nonlinear connections}
We assume now that our manifold is $\mathbb{R}^4$ endowed with standard coordinates
$(x^1,x^2,x^3,x^4=x,y,z,\xi=ct)$, and make some preliminary considerations in order to make the
choice of our projection $V$ consistent with the introduced concept of PhLO. The intrinsically
defined straight-line translational component of propagation of the PhLO will be assumed to be
parallel to the coordinate plane $(z,\xi)$. Also, $\frac{\partial}{\partial x}$ and
$\frac{\partial}{\partial y}$ will be vertical coordinate fields, so every vertical vector field
$Y$ can be represented by  $Y=f\,\frac{\partial}{\partial x}+g\,\frac{\partial}{\partial y}$. It is
easy to check that any two such linearly independent vertical vector fields $Y_1$ and $Y_2$ define
an integrable distribution, hence, the corresponding curvature will be zero. It seems very natural
to choose $Y_1$ and $Y_2$ to coincide correspondingly with the vertical projections of
$\frac{\partial}{\partial z}$ and $\frac{\partial}{\partial \xi}$. Moreover, let's restrict
ourselves to PhLO of electromagnetic nature and denote further the verical projection by $V$. Then,
since this vertical structure is meant to be smoothly straight-line translated along the plane
$(z,\xi)$ with the velocity of light, a natural suggestion comes to mind these two projections
$Y_1=V(\frac{\partial}{\partial z})$ and $Y_2=V(\frac{\partial}{\partial \xi})$) to be physically
interpreted as representatives of the electric and magnetic components. Now we know from classical
electrodynamics that the situation described corresponds to zero invariants of the electromagnetic
field, therefore, we may assume that $Y_1$ and $Y_2$ are ortogonal to each other and with the same
modules with respect to the euclidean metric in the 2-dimensional space spent by
$\frac{\partial}{\partial x}$ and $\frac{\partial}{\partial y}$. It follows that the essential
components of $Y_1$ and $Y_2$ should be expressible only with two independent functions $(u,p)$.
The conclusion is that our projection should depend only on $(u,p)$. Finally, we note that these
assumptions lead to the horizontal nature of $dz$ and $d\xi$.

Note that if the translational component of propagation is along the vector field $X$ then we can
define two new distributions : $(Y_1,X)$ and $(Y_2,X)$, which do not seem to be integrable in
general even if $X$ has constant components as it should be. Since these two distributions are
nontrivially intersected (they have a common member $X$), it is natural to consider them as
geometrical images of two consistently interacting physical subsystems of our PhLO. Hence, we must
introduce two projections with the same image space but with different kernal spaces, and the
components of both projections must depend only on the two functions $(u,p)$.

Let now $(u,p)$ be two smooth functions on $\mathbb{R}^4$ and
$\varepsilon=\pm 1$ . We introduce two projections $V$ and $\tilde{V}$ in
$T\mathbb{R}^4$ as follows:
\begin{equation}
V=dx\otimes\frac{\partial}{\partial
x}+dy\otimes\frac{\partial}{\partial y}
-\varepsilon\,u\,dz\otimes\frac{\partial}{\partial x}-
u\,d\xi\otimes\frac{\partial}{\partial x}-
\varepsilon\,p\,dz\otimes\frac{\partial}{\partial y}-
p\,d\xi\otimes\frac{\partial}{\partial y},
\end{equation}
\begin{equation}
\tilde{V}=
dx\otimes\frac{\partial}{\partial
x}+dy\otimes\frac{\partial}{\partial y} +p\,dz\otimes\frac{\partial}{\partial
x} +\varepsilon p\,d\xi\otimes\frac{\partial}{\partial x} -
u\,dz\otimes\frac{\partial}{\partial y}- \varepsilon
u\,d\xi\otimes\frac{\partial}{\partial y}.
\end{equation}
So, in both cases we consider $(\frac{\partial}{\partial x},\frac{\partial}{\partial y})$ as
vertical vector fields, and $(dz,d\xi)$ as horizontal 1-forms. By corresponding transpositions we
can determine projections $V^*$ and $\tilde{V}^*$ in the cotangent bundle $T^*\mathbb{R}^4$.
\[
V^*=dx\otimes\frac{\partial}{\partial x}+ dy\otimes\frac{\partial}{\partial y}
-\varepsilon\,u\,dx\otimes\frac{\partial}{\partial z}- u\,dx\otimes\frac{\partial}{\partial \xi}-
\varepsilon\,p\,dy\otimes\frac{\partial}{\partial z}- p\,dy\otimes\frac{\partial}{\partial \xi}, \]
\[ \tilde{V}^*= dx\otimes\frac{\partial}{\partial x}+ dy\otimes\frac{\partial}{\partial y}+
p\,dx\otimes\frac{\partial}{\partial z} + \varepsilon p\,dx\otimes\frac{\partial}{\partial \xi} -
u\,dy\otimes\frac{\partial}{\partial z}- \varepsilon u\,dy\otimes\frac{\partial}{\partial \xi}.
\]
The corresponding horizontal projections, denoted by $(H,H^*;\tilde{H},\tilde{H}^*)$ look as
follows:
\[ H=dz\otimes\frac{\partial}{\partial z}+d\xi\otimes\frac{\partial}{\partial \xi}
+\varepsilon\,u\,dz\otimes\frac{\partial}{\partial x}+ u\,d\xi\otimes\frac{\partial}{\partial x}+
\varepsilon\,p\,dz\otimes\frac{\partial}{\partial y}+ p\,d\xi\otimes\frac{\partial}{\partial y}, \]
\[ \tilde{H}=dz\otimes\frac{\partial}{\partial z}+d\xi\otimes\frac{\partial}{\partial \xi}-
p\,dz\otimes\frac{\partial}{\partial x} - \varepsilon p\,d\xi\otimes\frac{\partial}{\partial x}+
u\,dz\otimes\frac{\partial}{\partial y}+ \varepsilon u\,d\xi\otimes\frac{\partial}{\partial y}, \]
\[ H^*=dz\otimes\frac{\partial}{\partial z}+ d\xi\otimes\frac{\partial}{\partial \xi}
+\varepsilon\,u\,dx\otimes\frac{\partial}{\partial z}+ u\,dx\otimes\frac{\partial}{\partial \xi}+
\varepsilon p\,dy\otimes\frac{\partial}{\partial z}+ p\,dy\otimes\frac{\partial}{\partial \xi}, \]
\[ \tilde{H}^*=dz\otimes\frac{\partial}{\partial z}+ d\xi\otimes\frac{\partial}{\partial \xi}-
p\,dx\otimes\frac{\partial}{\partial z} - \varepsilon p\,dx\otimes\frac{\partial}{\partial \xi}+
u\,dy\otimes\frac{\partial}{\partial z}+ \varepsilon u\,dy\otimes\frac{\partial}{\partial \xi}. \]

The corresponding matrices look like:
\[ V=
\begin{Vmatrix}1 & 0 & -\varepsilon\,u & -u \\
0 & 1 & -\varepsilon\,p  & -p \\
0 & 0 & 0 & 0 \\
0 & 0 & 0 & 0 \end{Vmatrix} ,
\ \ H=
\begin{Vmatrix}0 & 0 & \varepsilon\,u & u \\
0 & 0 & \varepsilon\,p & p \\
0 & 0 & 1 & 0 \\
0 & 0 & 0 & 1 \end{Vmatrix} ,
\]

\[ V^*=
\begin{Vmatrix}1 & 0 & 0 & 0\\
0 & 1 & 0 & 0 \\
-\varepsilon\,u & -\varepsilon\,p & 0 & 0 \\
-u & -p & 0 & 0
\end{Vmatrix} ,
\ \ H^*=
\begin{Vmatrix}0 & 0 & 0 & 0\\
0 & 0 & 0 & 0 \\
\varepsilon\,u & \varepsilon\,p & 1 & 0 \\
u & p & 0 & 1 \end{Vmatrix} ,
\]

\vskip 0.4cm
\[
\tilde{V}= \begin{Vmatrix}1 & 0 & p & \varepsilon\,p \\
0 & 1 & -u & -\varepsilon\,u \\
0 & 0 & 0 & 0 \\
0 & 0 & 0 & 0 \end{Vmatrix} ,
\ \
\tilde{H}= \begin{Vmatrix}
0 & 0 & -p & -\varepsilon\,p \\
0 & 0 & u & \varepsilon\,u \\
0 & 0 & 1 & 0 \\
0 & 0 & 0 & 1 \end{Vmatrix} ,
\]
\vskip 0.4cm
\[
\tilde{V}^*= \begin{Vmatrix}1 & 0 & 0 & 0 \\
0 & 1 & 0 & 0\\
p & -u & 0 & 0 \\
\varepsilon\,p & -\varepsilon\,u & 0 & 0 \end{Vmatrix} ,
\ \
\tilde{H}^*= \begin{Vmatrix}
0 & 0 & 0 & 0 \\
0 & 0 & 0 & 0 \\
-p & u & 1 & 0 \\
-\varepsilon\,p & \varepsilon\,u & 0 & 1 \end{Vmatrix} .
\]
The
projections of the coordinate bases are:
\[
\left(\frac{\partial}{\partial x},
\frac{\partial}{\partial y}, \frac{\partial}{\partial z},
\frac{\partial}{\partial \xi}\right).V=
\left(\frac{\partial}{\partial x}, \frac{\partial}{\partial y},
-\varepsilon u\frac{\partial}{\partial x}
-\varepsilon p\frac{\partial}{\partial y},
-u\frac{\partial}{\partial x}
-p\frac{\partial}{\partial y}\right);
\]
\vskip 0.3cm
\[
\left(\frac{\partial}{\partial x},\frac{\partial}{\partial y},
\frac{\partial}{\partial z},\frac{\partial}{\partial \xi}\right).H=
\left(0,0,\varepsilon u\frac{\partial}{\partial x}
+\varepsilon p\frac{\partial}{\partial y}+\frac{\partial}{\partial z},
u\frac{\partial}{\partial x}
+p\frac{\partial}{\partial y}+
\frac{\partial}{\partial \xi}\right);
\]
%\newpage
%\vskip 0.3cm
\[
\left(dx,dy,dz,d\xi\right).V^*=
\left(dx-\varepsilon udz-ud\xi, dy-\varepsilon pdz-pd\xi,0,0\right)
\]
%\vskip 0.1cm
\[
\left(dx,dy,dz,d\xi\right).H^*=
\left(\varepsilon udz+ud\xi,\varepsilon pdz+pd\xi,dz,d\xi\right)
\]
\[
\left(\frac{\partial}{\partial x},
\frac{\partial}{\partial y}, \frac{\partial}{\partial z},
\frac{\partial}{\partial \xi}\right).\tilde{V}=
\left(\frac{\partial}{\partial x}, \frac{\partial}{\partial y},
p\frac{\partial}{\partial x}
-u\frac{\partial}{\partial y},
\varepsilon\,p\frac{\partial}{\partial x}
-\varepsilon\,u\frac{\partial}{\partial y}\right);
\]
\[
\left(\frac{\partial}{\partial x},\frac{\partial}{\partial y},
\frac{\partial}{\partial z},\frac{\partial}{\partial \xi}\right).\tilde{H}=
\left(0,0,-p\frac{\partial}{\partial x}
+u\frac{\partial}{\partial y}+\frac{\partial}{\partial z},
-\varepsilon\,p\frac{\partial}{\partial x}
+\varepsilon\,u\frac{\partial}{\partial y}+
\frac{\partial}{\partial \xi}\right);
\]
\[
\left(dx,dy,dz,d\xi\right).\tilde{V}^*=
\left(dx+p\,dz+\varepsilon\,pd\xi, dy-u\,dz-\varepsilon\,ud\xi,0,0\right)
\]
\[
\left(dz,d\xi,dx,dy\right).\tilde{H}^*=
\left(-p\,dz-\varepsilon\,p\,d\xi,u\,dz+\varepsilon\,u\,d\xi,dz,d\xi\right) .
\]
We compute now the two curvature 2-forms $\mathcal{R}$ and
$\tilde{\mathcal{R}}$. The components $\mathcal{R}^\sigma_{\mu\nu}$ of
$\mathcal{R}$ in coordinate basis are given by
$V^\sigma_\rho\Big(\big[H\frac{\partial}{\partial
x^{\mu}},H\frac{\partial}{\partial x^{\nu}}\big]^{\rho}\Big)$, and the only
nonzero components are just
$$
\mathcal{R}^{x}_{z\xi}=\mathcal{R}^{1}_{34}=
-\varepsilon(u_{\xi}-\varepsilon\,u_z),\ \ \
\mathcal{R}^{y}_{z\xi}=\mathcal{R}^{2}_{34}=
-\varepsilon(p_{\xi}-\varepsilon\,p_z)  .
$$
For the nonzero components of $\tilde{\mathcal{R}}$ we obtain
$$
\tilde{\mathcal{R}}^{x}_{z\xi}=\tilde{\mathcal{R}}^{1}_{34}=
(p_{\xi}-\varepsilon\,p_z),\ \ \
\tilde{\mathcal{R}}^{y}_{z\xi}=\tilde{\mathcal{R}}^{2}_{34}=
-(u_{\xi}-\varepsilon\,u_z)  .
$$

%Consider now the 2-forms:
%\[
%G=(P_V)^*dx\wedge (P_H)^*dx+(P_V)^*dy\wedge (P_H)^*dy=
%\]
%\[
%\varepsilon u\,dx\wedge dz+\varepsilon p\,dy\wedge dz+u\,dx\wedge d\xi+
%p\,dy\wedge d\xi
%\]
%\vskip 0.3cm
%\[
%\tilde{G}=
%(\tilde{P}_V)^*dx\wedge (\tilde{P}_H)^*dx+
%(\tilde{P}_V)^*dy\wedge (\tilde{P}_H)^*dy=
%\]
%\[
%-p\,dx\wedge dz+u\,dy\wedge dz-\varepsilon p\,dx\wedge d\xi+
%\varepsilon u\,dy\wedge d\xi
%\]
%\vskip 0.3cm
%\noindent
%It follows: $G=A\wedge\zeta, \ \tilde{G}=A^*\wedge\zeta$ and
%$\tilde{G}=*G$, where $*$ is the Hodge star operator defined by $\eta$.
%Clearly, the two 2-forms $(G,*G)$ represent the two nonintegrable Pfaff systems
%$(A,\zeta)$ and $(A^*,\zeta)$.

The corresponding two curvature forms are:
\begin{equation}
\mathcal{R}=-\varepsilon(u_\xi-\varepsilon u_z)dz\wedge d\xi\otimes
\frac{\partial}{\partial x}-
\varepsilon(p_\xi-\varepsilon p_z)dz\wedge d\xi\otimes
\frac{\partial}{\partial y}
\end{equation}
\begin{equation}
\mathcal{\tilde{R}}=(p_\xi-\varepsilon p_z)dz\wedge d\xi\otimes
\frac{\partial}{\partial x}-
(u_\xi-\varepsilon u_z)dz\wedge d\xi\otimes
\frac{\partial}{\partial y} .
\end{equation}
We obtain (in our coordinate system) $-\frac12tr\left(V\circ H^*\right)=
-\frac12tr\left(\tilde{V}\circ\tilde{H}^*\right)=u^2+p^2$, and
$$
V\left(\left[H\left(\frac{\partial}{\partial z}\right),
H\left(\frac{\partial}{\partial \xi}\right)\right]\right)=
\left[H\left(\frac{\partial}{\partial z}\right),
H\left(\frac{\partial}{\partial \xi}\right)\right]=
-\varepsilon(u_{\xi}-\varepsilon u_z)\frac{\partial}{\partial x}
-\varepsilon(p_{\xi}-\varepsilon p_z)\frac{\partial}{\partial y}\equiv Z_1,
$$
$$
\tilde{V}\left(\left[\tilde{H}\left(\frac{\partial}{\partial z}\right),
\tilde{H}\left(\frac{\partial}{\partial \xi}\right)\right]\right)=
\left[\tilde{H}\left(\frac{\partial}{\partial z}\right),
\tilde{H}\left(\frac{\partial}{\partial \xi}\right)\right]=
(p_{\xi}-\varepsilon p_z)\frac{\partial}{\partial x}
-(u_{\xi}-\varepsilon u_z)\frac{\partial}{\partial y}\equiv Z_2,
$$
where $Z_1$ and $Z_2$ coincide with the values of the two curvature forms
$\mathcal{R}$ and $\tilde{\mathcal{R}}$ on the coordinate vector fields
$\frac{\partial}{\partial z}$ and $\frac{\partial}{\partial \xi}$ respectively:
$$
Z_1=\mathcal{R}\left(\frac{\partial}{\partial z},
\frac{\partial}{\partial \xi}\right),\ \ \
Z_2=
\tilde{\mathcal{R}}\left(\frac{\partial}{\partial z},
\frac{\partial}{\partial \xi}\right) .
$$
We evaluate now the vertical 2-form
$V^*(dx)\wedge V^*(dy)$ on the bivector $Z_1\wedge Z_2$ and obtain
$\varepsilon\,\mathcal{K}^2$, where
$$
\mathcal{K}^2=(u_{\xi}-\varepsilon u_z)^2+(p_{\xi}-\varepsilon p_z)^2.
$$
An important parameter, having dimension
of length (the coordinates are assumed to have dimension of length)
and denoted by $l_o$, turns out to be the square rooth of the
quantity
$$ \frac{-\frac12tr\left(V\circ H^*\right)}{\mathcal{K}^2}=
\frac{u^2+p^2}{(u_{\xi}-\varepsilon
u_z)^2+(p_{\xi}-\varepsilon p_z)^2}.
$$
Clearly, if $l_o$ is finite constant it could be interpreted as some parameter
of extension of the PhLO described, so it could be used as identification
parameter in the dynamical equations and in lagrangians, but only if
$(u_{\xi}-\varepsilon u_z)\neq 0$ and $(p_{\xi}-\varepsilon p_z)\neq 0$. This
goes along with our concept of PhLO which does not admit spatially infinite
extensions.
Finally we'd like to note that the right-hand side of the above relation does
not depend on which projection $V$ or $\tilde{V}$ is used, i.e.
$[\tilde{V}^*(dx)\wedge\tilde{V}^*(dy)](Z_1\wedge Z_2)=
\varepsilon\,\mathcal{K}^2$ too, so
\begin{equation}
l_o^2=\frac{-\frac12tr\left(\tilde{V}\circ\tilde{H}^*\right)}{\mathcal{K}^2}=
\frac{-\frac12tr\Big(V\circ H^*\Big)}{\mathcal{K}^2}=
\frac{u^2+p^2}
{(u_{\xi}-\varepsilon u_z)^2+(p_{\xi}-\varepsilon p_z)^2}.
\end{equation}
The parameter $l_o$ has the following symmetry. Denote by
$V_o=dx\otimes\frac{\partial}{\partial x}+
dy\otimes\frac{\partial}{\partial y}$, then $V=V_o+V_1$ and
$\tilde {V}=V_o+\tilde{V}_1$, where, in our coordinates, $V_1$ and $\tilde
{V}_1$ can be seen above how they look like. We form now $W=aV_1-b\tilde{V}_1$
and $\tilde{W}=bV_1+a\tilde{V}_1$, where $(a,b)$ are two arbitrary real
numbers. The components of the corresponding projections $P_W=V_o+W$ and
$P_{\tilde{W}}=V_o+\tilde{W}$ can be obtained through the substitutions:
$u\rightarrow(au+\varepsilon bp); \ p\rightarrow (\varepsilon bp-ap)$. Now
$-\frac12tr(V\circ H^*)$ transforms to $(a^2+b^2)(u^2+p^2)$ and $\mathcal{K}^2$
transforms to $(a^2+b^2)[(u_{\xi}-\varepsilon u_z)^2+(p_{\xi}-\varepsilon
p_z)^2]$, so, $l_o(V)=l_o(W)$. This corresponds in some sense to the dual
summetry of classical vacuum electrodynamics. We note finally that the squared
modules of the two curvature forms $|\mathcal{R}|^2$ and
$|\mathcal{\tilde{R}}|^2$ are equal to $(u_{\xi}-\varepsilon
u_z)^2+(p_{\xi}-\varepsilon p_z)^2$ in our coordinates, therefore, the nonzero
values of $|\mathcal{R}|^2$ and $|\mathcal{\tilde{R}}|^2$, as well as the
finite value of $l_o$ guarantee that the two functions $u$ and $p$ are NOT
plane running waves.

 \vskip 0.5cm \section{Connection to classical electrodynamics}

From formal point of view the relativistic formulation of classical
electrodynamics in vacuum ($\rho=0$) is based on the following assumptions. The
configuration space is the Minkowski space-time $M=(\mathbb{R},\eta)$ where
$\eta$ is the pseudometric with $sign(\eta)=(-,-,-,+)$ with the corresponding
volume 4-form $\omega_o=dx\wedge dy\wedge dz\wedge d\xi $ and Hodge star $*$
defined by $\alpha\wedge\beta=\eta(*\alpha,\beta)\omega_o$. The electromagnetic
filed is describe by two closed 2-forms $(F,*F):\mathbf{d}F=0, \
\mathbf{d}*F=0$. The physical characteristcs of the field are deduced from the
following stress-energy-momentum tensor field
\begin{equation}
T_{\mu}{^\nu}(F,*F)=-\frac12\big[F_{\mu\sigma}F^{\nu\sigma}+
(*F)_{\mu\sigma}(*F^{\nu\sigma})\big].
\end{equation}
In the non-vacuum case the allowed energy-momentum exchange with other physical
systems is given in general by the divergence
\begin{equation}
\nabla_\nu\,T_{\mu}^{\nu}
=\frac12 \Big[F^{\alpha\beta}(\mathbf{d}F)_{\alpha\beta\mu}
+(*F)^{\alpha\beta}(\mathbf{d}*F)_{\alpha\beta\mu}\Big] =
F_{\mu\nu}(\delta F)^\nu + (*F)_{\mu\nu}(\delta *F)^\nu,
\end{equation}
where $\delta=*\mathbf{d}*$ is the coderivative.
If the field is free: $\mathbf{d}F=0, \mathbf{d}*F=0$, this divergence is
obviously equal to zero on the vacuum solutions since its both terms are zero.
Therefore, energy-momentum exchange between the two component-fields $F$ and
$*F$, which should be expressed by the terms
$(*F)^{\alpha\beta}(\mathbf{d}F)_{\alpha\beta\mu}$ and
$F^{\alpha\beta}(\mathbf{d}*F)_{\alpha\beta\mu}$ is NOT allowed on the
solutions of $\mathbf{d}F=0, \mathbf{d}*F=0$. This shows that the
widely used 4-potential approach (even if two 4-potentials $A,A^*$ are
introduced so that $\mathbf{d}A=F, \ \mathbf{d}A^*=*F$ locally)
to these equations excludes any possibility to individualize two
energy-momentum exchanging time-stable subsystems of the field that are
mathematically represented by $F$ and $*F$.

On the contrary, our concept of PhLO does NOT exclude such two
physically interacting subsystems of the field to really exist, and therefore,
to be mathematically individualized. The intrinsically connected two projections
$V$ and $\tilde{V}$ and the corresponding two curvature forms give the
mathematical realization of this idea: $V$ and $\tilde{V}$ individualize the
two subsystems, and the corresponding two curvature 2-forms $\mathcal{R}$ and
$\mathcal{\tilde{R}}$ represent the instruments by means of which mutual
energy-momentum exchange between these two subsystems could be locally performed.
Moreover, as we already mentioned, the energy-momentum tensor for a PhLO
must satisfy the additional local isotropy (null) condition
$T_{\mu\nu}(F,*F)T^{\mu\nu}(F,*F)=0$.

So, we have to construct appropriate quantities and relations
having direct physical sense in terms of the introduced and considered two
projections $V$ and $\tilde{V}$. The above well established in electrodinamics
relations say that we need two 2-forms to begin with.

Recall that our coordinate 1-forms $dx$ nd $dy$ have the following vertical and
horizintal projections:
\[
 V^*(dx)=dx-\varepsilon u\,dz-u\,d\xi, \ \
H^*(dx)=\varepsilon u\,dz+u\,d\xi\ , \]
\[
 V^*(dy)=dy-\varepsilon
p\,dz-p\,d\xi, \ \ H^*(dy)=\varepsilon p\,dz+p\,d\xi .
\]
We form now the 2-forms
$V^*(dx)\wedge H^*(dx)$ and $V^*(dy)\wedge H^*(dy)$:
$$
V^*(dx)\wedge
H^*(dx)=\varepsilon\,u\,dx\wedge dz+u\,dx\wedge d\xi,
$$
$$ V^*(dy)\wedge
H^*(dy)=\varepsilon\,p\,dy\wedge dz+p\,dx\wedge d\xi .
$$
Summing up these last
two relations and denoting the sum by $F$ we obtain
\begin{equation}
F=\varepsilon\,u\,dx\wedge dz+u\,dx\wedge d\xi+
\varepsilon\,p\,dy\wedge dz+p\,dy\wedge d\xi .
\end{equation}
Doing the same steps with $\tilde{V}^*$ and $\tilde{H}^*$ we obtain
\begin{equation}
\tilde{F}=-p\,dx\wedge dz-\varepsilon\,p\,dx\wedge d\xi+
u\,dy\wedge dz+\varepsilon u\,dy\wedge d\xi .
\end{equation}
Noting that our definition of the Hodge star requires
$(*F)_{\mu\nu}=-\frac12\,\varepsilon_{\mu\nu}\,^{\sigma\rho}F_{\sigma\rho}$,
it is now easy to verify that $\tilde{F}=*F$. Moreover, introducing the
notations
$$
A=u\,dx+p\,dy, \ \ A^*=-\varepsilon\,p\,dx+\varepsilon\,u\,dy ,\ \
\zeta=\varepsilon\,dz+d\xi ,
$$
we can represent $F$ and $\tilde{F}$ in the form
$$
F=A\wedge \zeta , \ \ \tilde{F}=*F=A^*\wedge \zeta .
$$
From these last relations
we see that $F$ and $*F$ are isotropic: $F\wedge F=0, F\wedge *F=0$, i.e. the
field $(F,*F)$ has zero invariants:
$F_{\mu\nu}F^{\mu\nu}=F_{\mu\nu}(*F)^{\mu\nu}=0$. The following relations are
now easy to verify:
\begin{equation}
V^*(F)=H^*(F)=V^*(*F)=H^*(*F)=\tilde{V}^*(F)=\tilde{H}^*(F)=
\tilde{V}^*(*F)=\tilde{H}^*(*F)=0,
\end{equation}
i.e. $F$ and $*F$ have zero vertical and horizontal projections with respect
to $V$ and $\tilde{V}$. Since, obviously, $\zeta $ is horizontal with respect
to $V$ and $\tilde{V}$ it is interesting to note that $A$ is vertical with
respect to $\tilde{V}$ and $A^*$ is vertical with respect to $V$:
$\tilde{V}^*(A)=A$, $V(A^*)=A^*$. In fact, for example, $$
\tilde{V}^*(A)=\tilde{V}^*(u\,dx+p\,dy)=u\tilde{V}^*(dx)+p\tilde{V}^*(dy)=
$$
$$
u[dx+p\,dz+\varepsilon p\,d\xi]+p[dy-u\,dz-\varepsilon u\,d\xi]=
u\,dx+p\,dy.
$$

We are going to establish now that there is real energy-momentum exchange
between the $F$-component and the $*F$-component of the field. To come to this
we compute the quantities $i(Z_1)F,\ \ i(Z_2)*F,\ \ i(Z_1)*F,\ \ i(Z_2)F$.
We obtain:
\[
i(Z_1)F=i(Z_2)*F=\langle A,Z_1\rangle\zeta=\langle A^*,Z_2\rangle\zeta=
\frac12\big[(u^2+p^2)_{\xi}-\varepsilon\,(u^2+p^2)_{z}\big]\zeta=
\]
\begin{equation}
=\frac12F^{\sigma\rho}(\mathbf{d}F)_{\sigma\rho\mu}dx^{\mu}=
\frac12(*F)^{\sigma\rho}(\mathbf{d}*F)_{\sigma\rho\mu}dx^{\mu}
=\frac12\nabla_\nu\,T_{\mu}^{\nu}(F,*F) ,
\end{equation}
\[
i(Z_1)*F=-i(Z_2)F=\langle A^*,Z_1\rangle\zeta =-\langle A,Z_2\rangle\zeta=
\big[u(p_{\xi}-\varepsilon\,p_{z})-p(u_{\xi}-\varepsilon\,u_{z})\big]\zeta=
\]
\begin{equation}
=-\frac12F^{\sigma\rho}(\mathbf{d}*F)_{\sigma\rho\mu}dx^{\mu}=
\frac12(*F)^{\sigma\rho}(\mathbf{d}F)_{\sigma\rho\mu}dx^{\mu} .
\end{equation}
If our field is free then $\nabla_\nu\,T_{\mu}^{\nu}(F,\tilde{F})=0$.
Moreover, in view of the divergence of the stress-energy-momentum tensor given
above, these last relations show that some real energy-momentum exchange between
$F$ and $*F$ takes place: the magnitude of the energy-momentum, transferred from
$F$ to $*F$ and given by
$i(Z_1)*F=\frac12(*F)^{\sigma\rho}(\mathbf{d}F)_{\sigma\rho\mu}dx^{\mu}$,
is equal to that, transferred from $*F$ to $F$, which is
given by
$-i(Z_2)F=-\frac12F^{\sigma\rho}(\mathbf{d}*F)_{\sigma\rho\mu}dx^{\mu}$.
On the other hand, as it is well known, the $*$-invariance
of the stress-energy-momentum tensor in case of zero invariants leads to
$F_{\mu\sigma}F^{\nu\sigma}=(*F)_{\mu\sigma}(*F)^{\nu\sigma}$, so, $F$ and $*F$
carry equal quantities of stress-energy-momentum. Physically this could mean
that the electromagnetic PhLO exist through a special internal dynamical
equilibrium between the two subsystems of the field, represented by $F$ and
$*F$, as mentioned in Section 1, namely, both subsystems carry the same
stress-energy-momentum and the mutual energy-momentum exchange between them is
always in equal quantities. This individualization does NOT mean that any of
the the two subsystems can exist separately, independently on each other.
Moreover, NO spatial "part" of PhLO is considered to represent a physical
object and to be energy-momentum carrier, as it is assumed, for example, when
mass and charge distributions are defined in classical electrodynamics.

\section{Equations of motion for electromagnetic PhLO}
Every system of equations describing the time-evolution of some physical system
should be consistent with the very system in the sense that all identification
characteristics of the system described must not change. In the case of
electromagnetic PhLO we assume the couple $(F,\tilde{F})$ to represent the
field, and in accordance with our notion for PhLO one of the identification
characteristics is straight-line translational propagation of the
energy-density with constant velocity "$c$", therefore, with every PhLO we may
associate appropriate direction, i.e. a geodesic null vector field
$X, X^2=0$ on the Minkowski space-time.
We choose further
$X=-\varepsilon \frac{\partial}{\partial z}+\frac{\partial}{\partial \xi}$,
which means that we have chosen the
coordinate sytem in such a way that the translational propagation is parallel to
the plane $(z,\xi)$. For another such parameter we assume that
the finite longitudinal extension of any PhLO is fixed and is given by an
appropriate positive number $\lambda $. In accordance with the "consistent
translational-rotational dynamical structure" of PhLO we shall assume that {\it
no translation is possible without rotation, and no rotation is possible
without translation}, and in view of the constancy of the translational
component of propagation we shall assume that the rotational component of
propagation is periodic, i.e. it is characterized by a constant frequency. The
natural period $T$ suggested is obviously $T=\frac{\lambda}{c}$. An obvoius
candidate for "rotational operator" is the linear map $J$ transforming $F$ to
$\tilde{F}$, which map coincides with the reduced to 2-forms Hodge-$*$, it
rotates the 2-frame $(A,A^*)$ to $\frac\pi 2$, so if such a rotation is
associated with a translational advancement of $l_o$, then a full rotation
should correspond to translational advancement of $4l_o=\lambda $. The simplest
and most natural translational change of the field $(F,\tilde{F})$
along $X$ should be given by
the Lie derivative of the field along $X$. Hence, the simplest and most natural
equations should read
\begin{equation}
\kappa l_o\,L_{X}(F)=\varepsilon\tilde{F},
\end{equation}
where
$F$ and $\tilde{F}$ are given in the preceding section,
$\kappa=\pm 1$ is responsible for left/right orientation of the rotational
component of propagation, and $l_o=const$. Vice versa, since $J\circ J=-id$ and
$J^{-1}=-J$ the above equation is equivalent to \[ \kappa
l_o\,L_{X}(\tilde{F})=-\varepsilon F. \] It is easy to show that these
equations are equivalent to
\begin{equation}
\kappa l_o\,L_{X}(V-V_o)=\varepsilon(\tilde{V}-V_o),
\end{equation}
where
$V_o=dx\otimes\frac{\partial}{\partial x}+
dy\otimes\frac{\partial}{\partial y}$
in our coordinates is the identity map in $Im(V)=Im(\tilde{V})$.
Another equivalent form is given by
$$
\kappa l_oZ_1=\bar{A^*}, \ \ \ \ \text{or} \ \ \ \  \kappa l_oZ_2=-\bar{A},
$$
where
$\bar{A^*}$ and $\bar{A}$ are $\eta$-corresponding vector fields to the 1-forms
$A^*$ and $A$.

Appropriate lagrangian for these equations ($l_o$=const.) is
\begin{equation}
\mathbb{L}=-\frac12\left(
\kappa l_o X^\sigma
\frac{\partial F_{\alpha\beta}}{\partial x^\sigma}-
\varepsilon \tilde{F}_{\alpha\beta}\right)
\tilde{F}^{\alpha\beta}+
\frac12\left(
\kappa l_o X^\sigma                          %4%
\frac{\partial \tilde{F}_{\alpha\beta}}{\partial x^\sigma}+
\varepsilon F_{\alpha\beta}\right)
F^{\alpha\beta},
\end{equation}
where $F$ and $\tilde{F}$, are considered as independent. The corresponding
Lagrange equations read
\begin{equation}
\kappa l_o X^\sigma                          %4%
\frac{\partial \tilde{F}_{\alpha\beta}}{\partial x^\sigma}+
\varepsilon F_{\alpha\beta}=0,\ \
\kappa l_o X^\sigma
\frac{\partial F_{\alpha\beta}}{\partial x^\sigma}-
\varepsilon \tilde{F}_{\alpha\beta}=0 .
\end{equation}
The stress-energy-momentum tensor is given by (13) under the additional
condition $T_{\mu\nu}T^{\mu\nu}=0$. It deserves noting that this isotropy
condition leads to zero invariants:
$$
I_1=F_{\mu\nu}F^{\mu\nu}=0, \ \  I_2=F_{\mu\nu}(*F)^{\mu\nu}=0, \ \
\text{and to} \ \ \
F_{\mu\sigma}F^{\nu\sigma}=(*F)_{\mu\sigma}(*F)^{\nu\sigma}.
$$
Hence, the two
subsystems represented by $F$ and $*F$ carry the same stress-energy-momentum,
therefore, $F\rightleftarrows *F$ energy-momentum exchange is possible only in
equal quantities.

In our coordinates the above equations reduce to
\[
\kappa l_o(u_{\xi}-\varepsilon\,u_z)=-p,\ \ \
\kappa l_o(p_{\xi}-\varepsilon\,p_z)=u,
\]
it is seen that the constant $l_o$ satisfies the above given relation (12).

From these last equations we readily obtain the relations
\[
(u^2+p^2)_{\xi}-\varepsilon\,(u^2+p^2)_z=0, \ \
u\,(p_{\xi}-\varepsilon\,p_z)-p\,(u_{\xi}-\varepsilon\,u_z)=
\frac{\kappa}{l_o}(u^2+p^2).
\]
Now, the substitution $u=\Phi\cos\,\psi,\ \  p=\Phi\sin\,\psi $, leads to the
relations
\[
L_{X}\Phi=0, \ \ L_{X}\psi=\frac{\kappa}{l_o}.
\]
Recalling now that $\Phi^2=-\frac12tr(V\circ H^*)$ and computing
$\frac12tr(V\circ L_{X}\tilde{H}^*)=
\varepsilon\big[u\,(p_{\xi}-\varepsilon\,p_z)-p\,(u_{\xi}-\varepsilon\,u_z)\big]=
\Phi^2\varepsilon\,L_{X}\psi$
the last two relations can be equivalently written as
\[
L_{X}\big[tr(V\circ H^*)\big]=0,\ \
tr(V\circ L_{X}\tilde{H}^*)=
-\frac{\varepsilon\,\kappa}{l_o}tr(V\circ H^*).
\]

It seems important to note the following. Another natural equations appear to
be the vacuum equations of Extended Electrodynamics (Donev,Tashkova 1995)
describing the internal energy-momentum redistribution during
evolution, namely,
\[
i(F)\mathbf{d}F=0,\ \ i(\tilde{F})\mathbf{d}\tilde{F}=0, \ \
i(\tilde{F})\mathbf{d}F=-i(F)\mathbf{d}\tilde{F}
\]
The class of nonlinear solutions to these equations, i.e. those satisfying
$\mathbf{d}\tilde{F}\neq 0,\ \  \mathbf{d}F\neq 0$, incorporates the solutions
to (23), however, at these conditions we obtain only one equation, namely,
\[
L_{X}\Phi^2=(u^2+p^2)_{\xi}-\varepsilon\,(u^2+p^2)_z=0, \ \
\]
which gives the energy conservation.
\vskip 2 cm
%			BEGINNING OF SECTION~\ref{Sect4
\section{Another look at the translational-rotational consistency}

In order to look at the translational-rotational consistency from a
point of view mentioned in the previous section  we recall the concept of
local symmetry of a distribution: a vector field $Y$ is a local (or
infinitesimal) symmetry of a p-dimensional distribution $\Delta$ defined by the
vector fields $(Y_1,\dots,Y_p)$ if every Lie bracket $[Y_i,Y]$ is in $\Delta$:
$[Y_i,Y]\in\Delta $. Clearly, if $\Delta$ is completely integrable, then
every $Y_i$ is a symmetry of $\Delta$, and the
flows of these vector fields move the points of each completely
integral manifold of $\Delta$ inside this completely integral manifold, that's
why they are called sometimes internal symmetries. If $Y$ is outside $\Delta$
then it is called {\it shuffling} symmetry, and in such a case the flow of $Y$
transforms a given completely integral manifold to another one. We are going to
show that our vector field
$X=-\varepsilon \frac{\partial}{\partial z}+\frac{\partial}{\partial \xi}$
is a shuffling symmetry for the distribution $\Delta_o$ defined by the vector
fields $(\bar{A},\bar{A}^*)$. But $\Delta_o$ coincides with our vertical
distribution generated by
$(\frac{\partial}{\partial x}, \frac{\partial}{\partial y})$, so it is
completely integrable and its integral manifold coincides with the
$(x,y)$-plane. From physical point of view this is expectable because the allowed
translational propagation of our PhLO along null straight lines should not
destroy it: this propagation just transforms the 2-plane $(x,y)$
passing through the point $(z_1,\xi_1)$ to a paralell to it 2-plane passing
through the point $(z_2,\xi_2)$, and these two points lay on the same
trajectory of our field $X$.

The corresponding Lie brackets are
\[
[\bar{A},X]=(u_\xi-\varepsilon\,u_z)\frac{\partial}{\partial x}+
(p_\xi-\varepsilon\,p_z)\frac{\partial}{\partial y}, \ \ \
[\bar{A}^*,X]=-\varepsilon\,(p_\xi-\varepsilon\,p_z)\frac{\partial}{\partial x}+
\varepsilon\,(u_\xi-\varepsilon\,u_z)\frac{\partial}{\partial y}.
\]
We see that $[\bar{A},X]$ and $[\bar{A}^*,X]$ are generated by
$(\frac{\partial}{\partial x}, \frac{\partial}{\partial y})$,
but $X$ is outside $\Delta_o$, so our field $X$ is a
shuffling local symmetry of $\Delta_o$.

We notice now that at each point our projections $V$ and $\tilde{V}$ generate
two frames:
$(\bar{A},\bar{A}^*,\partial_z,\partial_{\xi})$ and
$([\bar{A},X],[\bar{A^*},X],\partial_z,\partial_{\xi})$.
Physically this would mean that the internal
energy-momentum redistribution during propagation
transforms the first frame into the second one and vice versa,
since both are defined by the dynamical nature of our PhLO. Taking into
account that only the first two vectors of these two frames change during
propagation we write down the corresponding linear transformation as follows:
\[
([\bar{A},X],[\bar{A^*},X])=(\bar{A},\bar{A}^*)
\begin{Vmatrix}\alpha & \beta \\ \gamma & \delta\end{Vmatrix} .
\]
Solving this system with respect to the real
numbers $(\alpha,\beta,\gamma,\delta)$ we obtain
\[
\begin{Vmatrix}\alpha & \beta \\ \gamma & \delta \end{Vmatrix}=
\frac{1}{\phi^2}
\begin{Vmatrix} -\frac12 L_{X}\Phi^2 &
\varepsilon \mathbf{R} \\ -\varepsilon \mathbf{R}
& -\frac12 L_{X}\Phi^2 \end{Vmatrix}=
-\frac12\frac{L_{X}\Phi^2}{\Phi^2}
\begin{Vmatrix} 1 & 0 \\ 0 & 1\end{Vmatrix}+
\varepsilon L_{X}\psi\begin{Vmatrix} 0 & 1 \\ -1 & 0\end{Vmatrix} ,
\]
where $\mathbf{R}=u\,(p_\xi-\varepsilon\,p_z)-p\,(u_\xi-\varepsilon\,u_z)$.
If the translational propagation is governed by the conservation law
$L_{X}\Phi^2=0$, then we obtain that the rotational component of
propagation is governed by the matrix $\varepsilon L_{X}\psi\,J$, where $J$
denotes the canonical complex structure in $\mathbb{R}^2$, and since
$\Phi^2\,L_{X}\psi=u\,(p_\xi-\varepsilon\,p_z)-p\,(u_\xi-\varepsilon\,u_z)\neq
0$ we conclude that the rotational component of propagation would be available
if and only if $\mathbf{R}\neq 0$. We may also say that a consistent
translational-rotational dynamical structure is available if the amplitude
$\Phi^2=u^2+p^2$ is a running wave along $X$ and the phase
$\psi=\mathrm{arctg}\frac{p}{u}$ is NOT a running wave along $X : L_{X}\psi\neq
0$. Physically this means that the rotational component of propagation is
entirely determined by the available internal energy-momentum exchange:
$i(\tilde{F})\mathbf{d}F=-i(F)\mathbf{d}\tilde{F}$.

Now if we have to
{\it guarantee the
conservative and constant character of the rotational aspect} of the PhLO
nature, we can assume
$L_{X}\psi=const=\kappa l_o^{-1}, \kappa=\pm 1$.
Thus, the frame rotation
$(\bar{A},\bar{A^*},\partial_z, \partial_\xi)\rightarrow
([\bar{A},X],[\bar{A^*},X],\partial_z, \partial_\xi)$,
i.e. $[\bar{A},X]=-\varepsilon\bar{A^*}\,L_{X}\psi$
and $[\bar{A^*},X]=\varepsilon\bar{A}\,L_{X}\psi $,
gives the following equations for the two functions $(u,p)$:
\[
u_\xi-\varepsilon u_z=-\frac{\kappa}{l_o}\,p, \ \ \
p_\xi-\varepsilon p_z=\frac{\kappa}{l_o}\,u \ .
\]
If we now quite independently from the projections considered introduce the
complex valued function $\Psi=u\,I+p\,J$, where $I$ is the identity map in
$\mathbb{R}^2$, the above two equations are formally equivalent to $$
L_{X}\Psi=\frac{\kappa}{l_o}J(\Psi) \ ,
$$
which appearantly demonstrates the translational-rotational consistency
in the above declared form that {\it no translation is possible without
rotation, and no rotation is possible without translation}, where the rotation
is represented by the complex structure $J$.

The quantity
$\mathbf{R}=u\,(p_\xi-\varepsilon\,p_z)-p\,(u_\xi-\varepsilon\,u_z)=
\Phi^2L_{X}\psi=\kappa l_o^{-1}\Phi^2$ suggests
to find an integral characteristic of the PhLO rotational nature. In fact, the
two co-distributions $(A,\zeta)$ and $(A^*,\zeta)$ define the two (equal in our
case) Frobenius 4-forms
$\mathbf{d}A\wedge A\wedge \zeta=\mathbf{d}A^*\wedge A^*\wedge \zeta$.
Each of these two 4-forms is equal to $\varepsilon\mathbf{R}\omega_o$.
Now, multiplying by $l_o/c$ any of them we obtain:
\begin{equation}
\frac{l_o}{c}\,\mathbf{d}A\wedge A\wedge \zeta=
\frac{l_o}{c}\,\mathbf{d}A^*\wedge A^*\wedge \zeta=
\frac{l_o}{c}\varepsilon\mathbf{R}\omega_o =
\varepsilon\kappa\frac{\Phi^2}{c}\omega_o\ .
\end{equation}
Integrating over the 4-volume $\mathbb{R}^3\times(\lambda=4l_o)$ (and having in
view the spatially finite nature of PhLO) we obtain the finite quantity
$\mathcal{H}=\varepsilon\kappa ET$, where $E$ is the integral energy of
the PhLO, $T=\frac{\lambda}{c}$, which clearly is the analog of the Planck
formula $E=h\nu$, i.e. $h=ET$. The combination $\varepsilon\kappa$ means that
the two orientations of the rotation, defined by $\kappa=\pm 1$, may be
observed in each of the two spatial directions of translational propagation
of the PhLO along the $z$-axis: from $-\infty$ to  $+\infty$, or from $+\infty$
to $-\infty$.
\vskip 0.5cm
\section{Solutions}
We consider the equations obtained in terms of the two functions
$\Phi=\sqrt{u^2+p^2}$ and $\psi=arctg\frac{p}{u}$. The equation for $\Phi$
in our coordinates is $\Phi_{\xi}-\varepsilon \Phi_{z}=0$, therefore,
$\Phi=\Phi(x,y,\xi+\varepsilon z)$. The equation for $\psi
$ is $\psi_{\xi}-\varepsilon\psi_{z}=\frac{\kappa}{l_o}$. Two families of
solutions for $\psi$, depending on an arbitrary function $\varphi$ can be given
by
\[
\psi_1=-\frac{\varepsilon\kappa}{l_o}z+\varphi(x,y,\xi+\varepsilon z),\ \ \
\text{and}\ \ \ \psi_2=\frac{\kappa}{l_o}\xi+\varphi(x,y,\xi+\varepsilon z) .
\]
Since $\Phi^2$ is a spatially finite function representing the energy density
we see that the translational propagation of our PhLO is represented by a
spatially finite running wave along the $z$-coordinate. Let's assume that the
phase is given by $\psi_1$ and $\varphi=const$. The form of this solution
suggests to choose the initial condition
$u_{t=0}(x,y,\varepsilon z),p_{t=0}(x,y,\varepsilon z)$ in the
following way. Let for $z=0$ the initial condition be
located on a disk $D=D(x,y;a,b;r_o)$ of small radius $r_o$, the center of the
disk to have coordinates $(a,b)$, and the value of
$\Phi_{t=0}(x,y,0)=\sqrt{u_{t=0}^2+p_{t=0}^2}$ to be
proportional to some appropriate for the case bump function $f$ on $D$ of the
distance $\sqrt{(x-a)^2+(y-b)^2}$ between the origin of the coordinate system
and the point $(x,y,0)$, such that it is centered at the point $(a,b)$, so,
$f(x,y)=f(\sqrt{(x-a)^2+(y-b)^2})$, $D$ is defined by
$D=\{(x,y)|\sqrt{(x-a)^2+(y-b)^2}\leq r_o\}$, and $f(x,y)$ is zero outside $D$.
Let also the dependence of $\Phi_{t=0}$ on $z$ be given by be the corresponding
bump function $\theta(z;\lambda)$ of an interval $(z,z+\lambda)$ of length
$\lambda=4l_o$ on the $z$-axis. If $\gamma$ is the proportionality coefficient
we obtain
\begin{align*}
u=\gamma\,\Phi(x,y,z,ct+\varepsilon z;a,b,\lambda)\,
\theta(ct+\varepsilon z;\lambda)\,cos(\psi_1), \\
p=\gamma\,\Phi(x,y,z,ct+\varepsilon z;a,b,\lambda)\,
\theta(ct+\varepsilon z;\lambda)\,sin(\psi_1).
\end{align*}
We see that because of the available {\it sine} and {\it cosine} factors in
the solution, the initial condition for the solution will occupy a $3d$-spatial
region of shape that is close to a helical cylinder of height $\lambda$, having
internal radius of $r_o$ and wrapped up around the $z$-axis. Also, its center
will always be $\sqrt{a^2+b^2}$-distant from the $z$-axis. Hence, the solution
will propagate translationally along the coordinate $z$ with the velocity $c$,
and, rotationally, inside the corresponding infinitely long helical cylinder
because of the $z$-dependence of the available periodical multiples.

On the two figures below are given two theoretical examples with $\kappa=-1$
and $\kappa=1$ respectively, amplitude function $\Phi$ located inside a
one-step helical cylinder with height of $\lambda$, and phase \newline
$\psi=\kappa \frac{2\pi z}{\lambda}$. The solutions
propagate left-to-right, i.e. $\varepsilon=-1$, along the coordinate $z$.
\vskip0.5cm
\begin{center}
\begin{figure}[ht!]
\centerline{
{\mbox{\psfig{figure=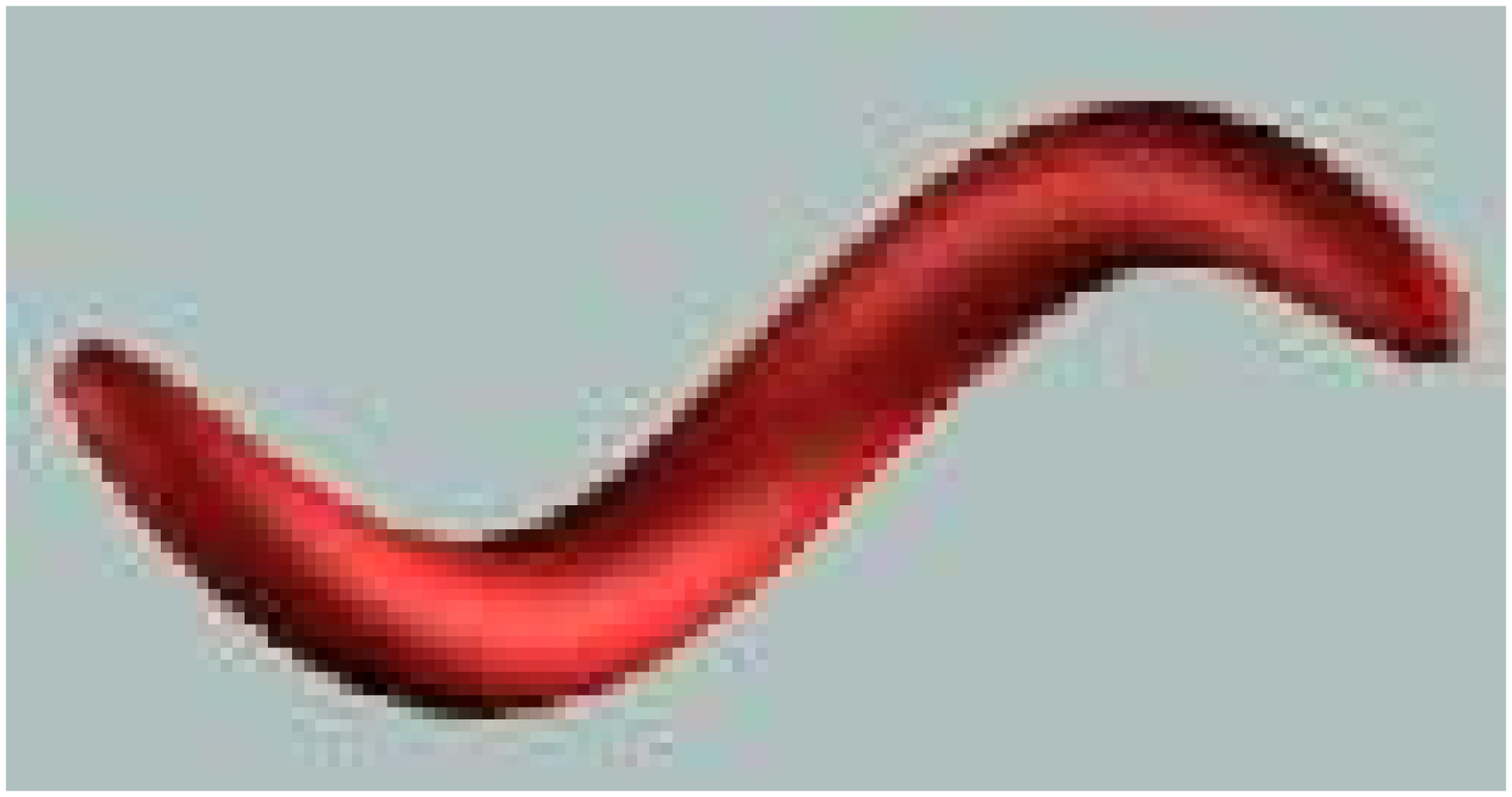,height=1.8cm,width=3.5cm}}
\mbox{\psfig{figure=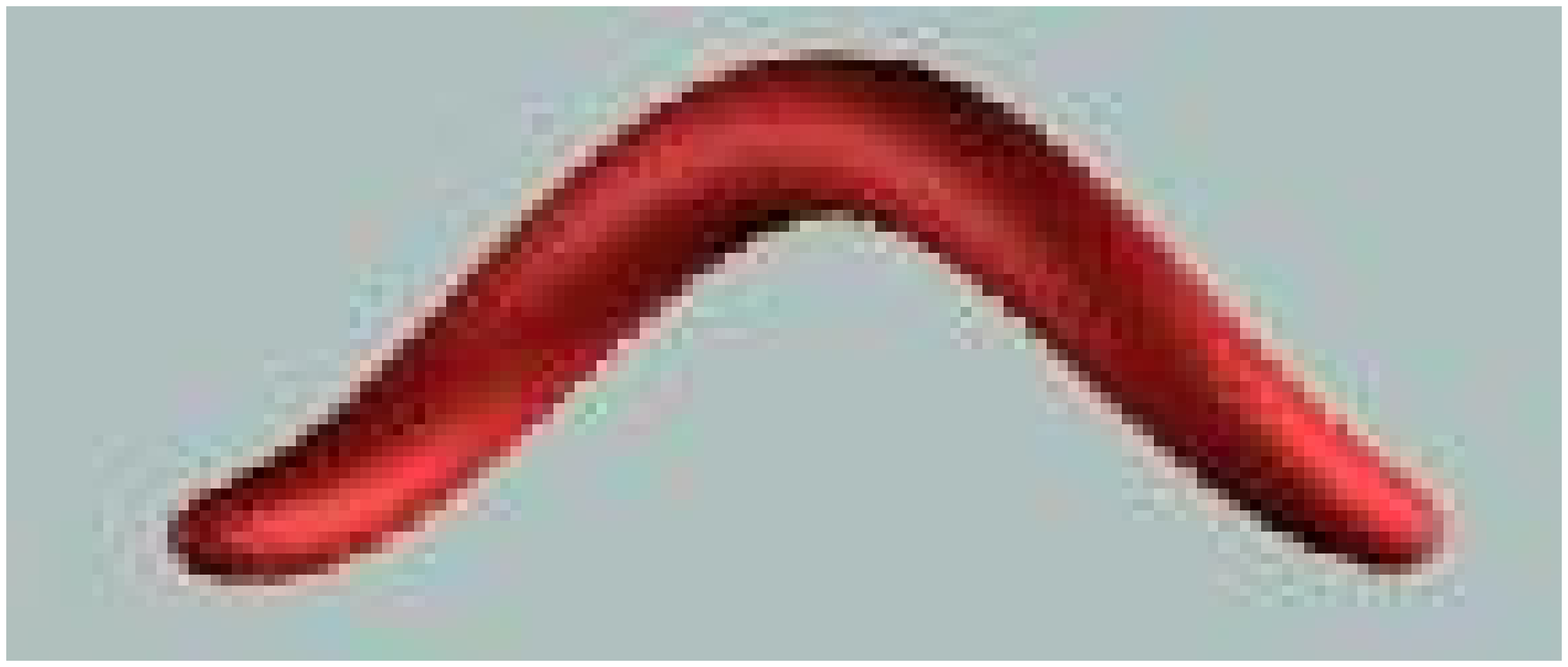,height=1.8cm,width=4.2cm}}
\mbox{\psfig{figure=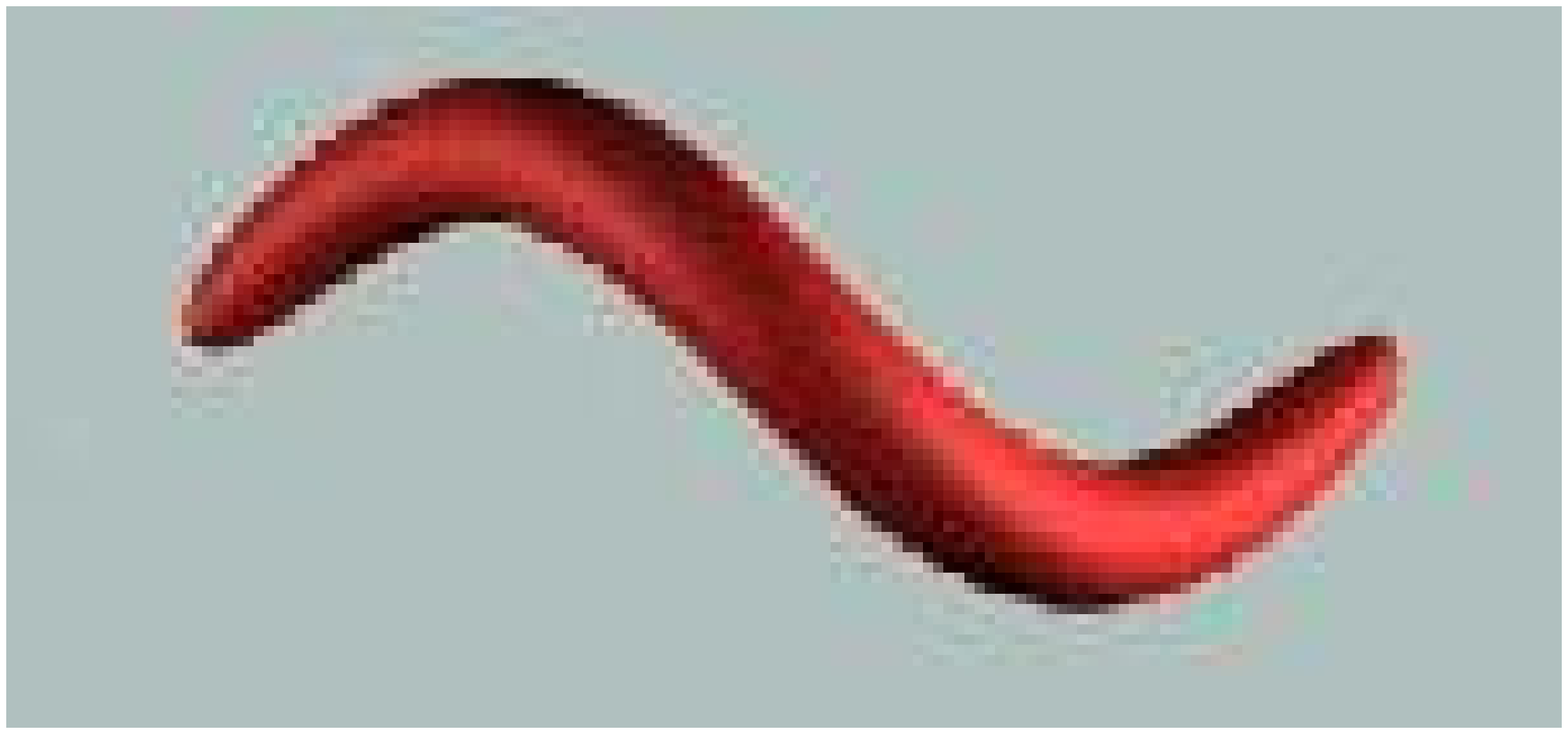,height=1.8cm,width=4.2cm}}}}
\caption{Theoretical example with $\kappa=-1$. The translational propagation is
directed left-to-right.}
\end{figure}
\end{center}
\begin{center}
\begin{figure}[ht!]
\centerline{
{\mbox{\psfig{figure=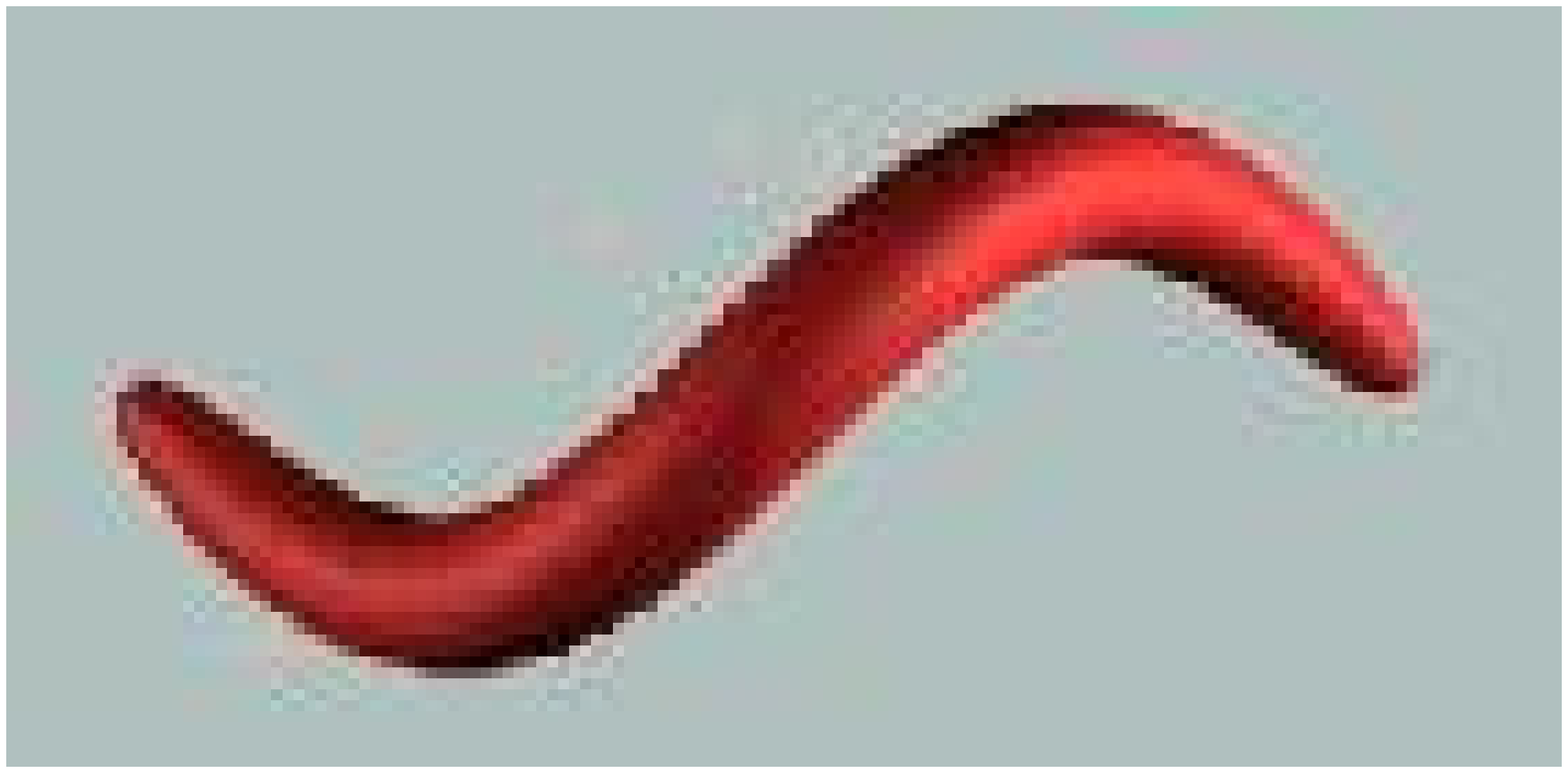,height=1.8cm,width=3.5cm}}
\mbox{\psfig{figure=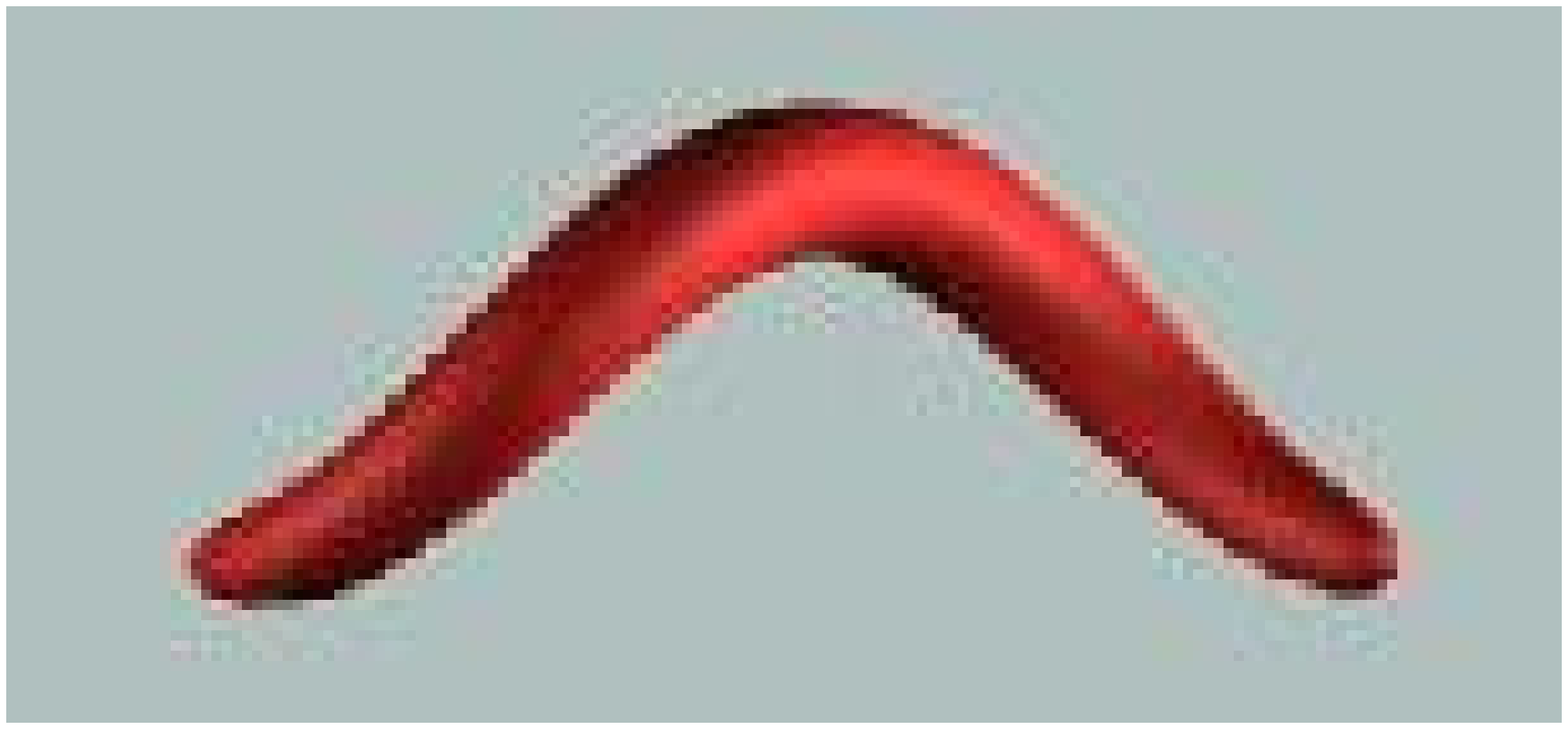,height=1.8cm,width=4.2cm}}
\mbox{\psfig{figure=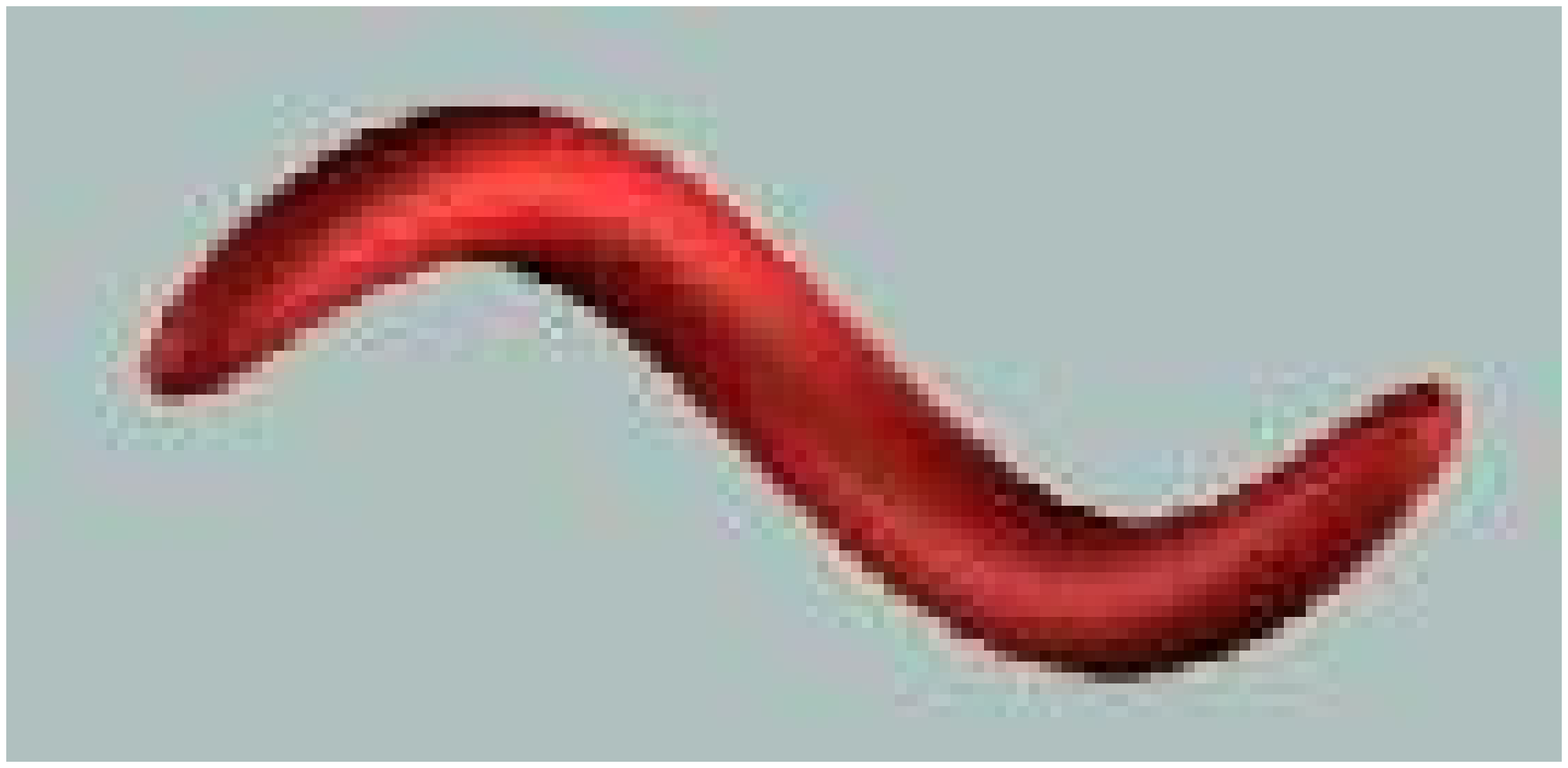,height=1.8cm,width=4.2cm}}}}
\caption{Theoretical example with $\kappa=1$. The translational propagation is
directed left-to-right.}
\end{figure}
\end{center}
%\vskip 0.5cm
The curvature $K$ and the torsion $T$ of the screwline through the
point $(x,y,0)\in D$ will be \[ K=\frac{R(x,y,0)}{R^2(x,y,0)+b^2},\ \ \ \
T=\frac{\kappa\,b}{R^2(x,y,0)+b^2} \ , \] where $b=\lambda/2\pi$. The
rotational frequency $\nu$ will be $\nu=c/2\pi b$, so we can introduce
period $T=1/\nu$ and elementary action $h=E.T$, where $E$ is the
(obviously finite) integral energy of the solution defined as 3d-integral
of the energy density $\Phi^2$.
\section{Conclusion}
We introduced a notion of PhLO as a
spatially finite physical object with a consistent translational-rotational
dynamical structure, and  built a corresponding mathematical model making use
of the geometry of nonintegrable distributions, i.e. nontrivial nonlinear
connections, on a manifold. This approach to PhLO we consider as an
illustration of the general idea that physical objects with dynamical structure
seem to be in a good, local as well as integral, correspondence with an
completely integrable distribution $\Delta$ plus an appropriate set $\Sigma$ of
nonintegrable subdistributions $\Delta_k, k=1,\dots,p\, ; p<n $ on a manifold
$M^n$, such that their curvature 2-forms $\Omega_k$ send couples of vector
fields from $\Delta_k\in\Sigma$ into $\Delta_m\in\Sigma$, where $k\neq m$, so
that, $\Omega_k(X,Y)\in\Sigma_m,\, (X,Y)\in \Delta_k$.

The two basic features of our approach to describe the dynamical structure
and behaviour of electromagnetic PhLO are: {\bf first, from physical viewpoint},
two dynamically interacting subsytems of a PhLO can be individualized, these
subsystems carry the same stress-energy-momentum, and they exchange
energy-momentum locally always in equal quantities, so they exist in a {\it
dynamical equilibrium}; {\bf second, from mathematical viewpoint}, to every
PhLO a couple of two nonlinear connections $V$ and $\tilde{V}$ is associated,
such that they have a common image space, and their inter-communication is
carried out and guaranteed by the two nonzero curvature forms $\Omega$ and
$\tilde{\Omega}$. The values of $\Omega$ and
$\tilde{\Omega}$ define two linearly independent 1-dimensional spaces, so, the
corresponding two exterior products with the direction of translational
propagation gives the mathematical images $F$ and $\tilde{F}$ of the two
subsystems. This approach allows to get some information concerning the
dynamical nature of the PhLO structure not only algebraically, i.e. only
through $V$ and $\tilde{V}$, but also {\it infinitesimally}, i.e. through the
curvature forms. While the energy density $\Phi^2$ of a PhLO propagates only
translationaly along straight isotropic lines, the available interaction of the
two subsytems of a PhLO demonstrates itself through a rotational component of
the entire propagational behaviour. The mutual energy-momentum exchanges are
given by the interior products of the images of $\Omega$ and $\tilde{\Omega}$
with $F$ and $\tilde{F}$, in particular, the dynamical equilibrium between $F$
and $\tilde{F}$ is given by
$i_{\tilde{F}}(\mathbf{d}F)=-i_{F}(\mathbf{d}\tilde{F})$.

Besides the spatially finite nature of PhLO that is allowed by our model and
illustrated with the invariant parameter $l_o$, two basic identifying
properties of PhLO were substantially used: straight-line translational
propagation with constant speed, and constant character of the rotational
component of propagation. The physical characteristics of a PhLO are
represented by an analog of the Maxwell-Minkowski stress-energy-momentum
tensor. An interesting moment is that $F$ and $\tilde{F}$ have zero horizonal
and vertical components with respect to the two nonlinear connections. The
equations of motion can be viewed from different viewpoints: as consistency
conditions between the rotational and translational components of propagation,
as Lagrange equations for an action principle, as part of the solutions of the
vacuum equations of extended electrodynamics, and also as naturally defined
transformation of 2-dimensional frames. In all these aspects of the equations
of motion the curvature forms play essential role through contralling the
inter-commumication between $F$ and $\tilde{F}$. Moreover, the Frobenius
curvature turns out to be proportional to the energy density, which allows an
analog of the famous Planck formula to be introduced. Moreover, this
"enrgy-density - curvature" proportionality goes along with the main
idea of General Relativity.

The solutions considered illustrate quite well the positive aspects of our
approach. It is interesting to note that the phase terms depend substantially
only on spatial variables, so, the spatial structure of the solutions
considered participates directly in the rotational component of the PhLO
dynamical structure.

The studies that resulted in writing this paper have been supported by Contract
$\phi 1515$ with the Bulgarian Svience Research Fund.

 \vskip 0.6cm
 {\bf References} \vskip 0.3cm Dainton,
J., 2000, {\it Phil. Trans. R. Soc. Lond. A}, {\bf 359}, 279 \vskip 0.3cm De
Broglie, L. V., 1923,  {\it Ondes et quanta}, C. R. {\bf 177}, 507 \vskip 0.3cm
Donev, S., Tashkova, M., 1995, {\it Proc. R. Soc. Lond. A} {\bf 450}, 281
(1995);

 arXiv:hep-th/0403244
\vskip 0.3cm
Einstein, A. 1905, {\it Ann. d. Phys.}, {\bf 17}, 132
\vskip 0.3cm
Godbillon, C., 1969,  {\it Geometrie differentielle et mecanique analytiqe},
Hermann, Paris
\vskip 0.3cm Godbole, R. M., 2003, arXiv: hep-th/0311188
\vskip 0.3cm
Lewis, G. N., 1926, {\it Nature}, {\bf 118}, 874
\vskip 0.3cm
Nisius, R., 2001,  arXiv: hep-ex/0110078
\vskip 0.3cm
Planck, M. 1901, {\it Ann. d. Phys.}, {\bf 4}, 553
\vskip 0.3cm
Speziali, P., 1972, Ed. {\it Albert Einstein-Michele Besso
Correspondence} (1903-1955),

Herman, Paris, 453
\vskip 0.3cm
Stumpf, H., Borne, T., 2001, {\it Annales de la Fond. Louis De Broglie},
{\bf 26}, No. {\it special}, 429
\vskip 0.3cm
Synge, J., 1958, {\it Relativity: the special theory}, Amsterdam, North Holland
\vskip 0.3cm
Vacaru, S. et al. {\it Clifford and Riemann-Finsler struvtures in geometric
mechanics an Gravity},

Selected works; arXiv:gr-qc/0508023v2.

\end{document}